\documentclass[a4paper,11pt]{article}

\usepackage[margin=1.14in]{geometry} 
\usepackage{tabularx}
\usepackage{multirow}
\usepackage{arydshln}
\usepackage{amsmath}
\usepackage{amsthm}
\usepackage{amssymb}
\usepackage{graphicx}
\usepackage{pdfpages}
\makeatletter
\makeatletter \renewcommand{\@dotsep}{10000} \makeatother
\usepackage{wrapfig}
\usepackage[utf8]{inputenc}
\usepackage{lmodern}
\usepackage{hyperref}
\hypersetup{
	unicode,
	colorlinks,
	breaklinks,
	urlcolor=cyan, 
    linkcolor=black, 
	pdfauthor={Author One, Author Two, Author Three},
	pdfproducer={LaTeX},
	pdfcreator={pdflatex}
}
\usepackage[T1]{fontenc} % if needed
\usepackage{amsmath}

\newcommand{\be}{\begin{eqnarray}}
\newcommand{\ee}{\end{eqnarray}}
\def\be{\begin{equation}}
\def\ee{\end{equation}}
\def\bea{\begin{eqnarray}}
\def\eea{\end{eqnarray}}

\newcommand{\gsim}{\;\raisebox{-0.9ex}{$\textstyle\stackrel{\textstyle >}{\sim}$}\;}
\newcommand{\lsim}{\;\raisebox{-0.9ex}{$\textstyle\stackrel{\textstyle<}{\sim}$}\;}
\def\lsim{\raise0.3ex\hbox{$\;<$\kern-0.75em\raise-1.1ex\hbox{$\sim\;$}}}
\def\gsim{\raise0.3ex\hbox{$\;>$\kern-0.75em\raise-1.1ex\hbox{$\sim\;$}}}

\usepackage{graphics}

\usepackage{epsfig}
\usepackage{slashed}
\usepackage[utf8]{inputenc}
\usepackage{multirow}
\usepackage{pstricks}
\usepackage{dcolumn}

\usepackage[sort&compress,numbers,square]{natbib}
\theoremstyle{plain}

\theoremstyle{definition}

\usepackage{graphicx, color}

\title{Multi-scale  cross-attention transformer encoder \\[0.25cm] for event classification}
\vspace{8mm}
\author{\Large{A. Hammad$^{a}$,  S. Moretti$^{b, c}$ and M. Nojiri$^{a,d,e}$}}

\date{
\small{
$^a$Theory Center, IPNS, KEK, 1-1 Oho, Tsukuba, Ibaraki 305-0801, Japan.\\
$^b$School of Physics and Astronomy, University of Southampton, Highfield,
Southampton, UK.\\
$^c$Department of Physics $\&$ Astronomy, Uppsala University, Box 516, SE-751 20 Uppsala, Sweden.\\
$^d$The Graduate University of Advanced Studies (Sokendai), 1-1 Oho, Tsukuba, Ibaraki 305-0801, Japan\\
$^e$Kavli IPMU (WPI), University of Tokyo, 5-1-5 Kashiwanoha, Kashiwa, Chiba 277-8583, Japan }
}
\begin{document}
	\maketitle
	\vspace{4mm}
	\begin{abstract}
 \normalsize{We deploy an advanced Machine Learning (ML) environment, leveraging a multi-scale cross-attention encoder for event classification, towards the identification of the $gg\to H\to hh\to b\bar b b\bar b$ process at the High Luminosity Large Hadron Collider (HL-LHC), where $h$ is the discovered Standard Model (SM)-like Higgs boson and $H$ a heavier version of it (with $m_H>2m_h$).  
 In the ensuing boosted Higgs regime, the final state consists of two fat jets. Our multi-modal network can extract information from the jet substructure and the kinematics of the final state particles through self-attention transformer layers. The diverse learned information is subsequently integrated to improve classification performance using an additional transformer encoder with cross-attention heads. We ultimately prove that our approach surpasses current alternative methods used to establish sensitivity to this process in performance, whether solely based on kinematic analysis or combining this with mainstream ML approaches. Then, we employ various interpretive methods to evaluate the network results, including attention map analysis and visual representation of Gradient-weighted Class Activation Mapping (Grad-CAM). Finally, we note that the proposed network is generic and can be applied to analyse any process carrying information at different scales.  Our code is publicly available for generic use\footnote{\href{https://github.com/AHamamd150/Multi-Scale-Transformer-Encoder}{https://github.com/AHamamd150/Multi-Scale-Transformer-Encoder.}}.
 }
\end{abstract}
\newpage
\noindent\rule{\textwidth}{1pt}
\tableofcontents
\noindent\rule{\textwidth}{0.2pt}
\maketitle \flushbottom
\vspace{4mm}
%%%%%%%%%%%%%%%%%%%%%%%%%%%%%%%%%%%%%%%%%%%%%%%%%%%%%%%%%%%%%%%%%%%%%%%%%%
\section{Introduction}
\label{sec:intro}
%%%----------------------%%%%%%I
Information about jet identification provides powerful insights into collision events and can help to separate different physics processes originating these. 
This information can be extracted from the elementary particles localized inside a jet. 
Recently, various methods have been used to exploit the substructure of a jet to probe new physics signatures using advanced Machine Learning (ML) techniques  \cite{Chakraborty:2019imr,Chung:2020ysf,Guo:2020vvt,Khosa:2021cyk,Datta:2019ndh}.

Conversely, using the reconstructed kinematics from the final state jets for event classification spans the full phase space and exhibits large classification performance \cite{Cogollo:2020afo,Grossi:2020orx,Ngairangbam:2020ksz,Englert:2020ntw,Freitas:2020ttd,Stakia:2021pvp,Jorge:2021vpo,Ren:2021prq,Alvestad:2021sje,Jung:2021tym,Drees:2021oew,Cornell:2021gut,
Vidal:2021oed}. Such high-level kinematics (i.e., encoding the global features of the final state particles), possibly together with the knowledge of the properties of (known or assumed) resonant intermediate particles, remains blind to the information encoded inside the final state jets.

A possible way to extract information from both jet substructure and global jet kinematics is to concatenate the information extracted from a multi-modal network \cite{Lin:2018cin,Moreno:2019neq,Chung:2022kjp,Kim:2019wns,Huang:2022rne,Esmail:2023axd}. However, such a simple concatenation leads to an imbalance of the extracted information, within which the kinematic information generally dominates \cite{Ban:2023jfo}.

In this paper, we present a novel method for incorporating different-scale information extracted from both global kinematics and substructure of jets via a transformer encoder with a cross-attention layer. The model initially extracts the most relevant information from each dataset individually using self-attention layers before incorporating these using a cross-attention layer. The method demonstrates a larger improvement in classification performance compared to the simple concatenation method.

To assess our results, we analyze the learned information by the transformer layers through the examination of the attention maps of the self- and cross-attention layers. 
Attention maps provide information about the (most) important embedded particles the model focuses on when classifying signal and background events. 
However, they cannot highlight the region in the feature (e.g., phase) space crucial for model classification. For this purpose, we utilize Gradient-weighted Class Activation Mapping (Grad-CAM) to highlight the geometric region in the $\eta-\phi$ (detector) plane where the model focuses on classifying events.

We test our approach for the dominant decay channel of Higgs boson pairs with Standard Model properties ($hh$) produced at the LHC, that is, into four $b$-(anti)quarks. 
%We do so as 
This signal has historically proved to be extremely challenging to extract owing to a significant QCD background. Lately, there have been several attempts to tackle this signature using both standard \cite{Chakraborty:2020vwj,Chakraborty:2022lcj,Chakraborty:2023hrk} andML \cite{Cerro:2022rpf} approaches. 
Furthermore, in the case that the $hh$ intermediate state emerges from the (resonant) decay of a heavier Higgs state ($H$), the kinematics becomes very challenging, as each of the two would-be (slim) $b$-jets produced by the two $h$ decays actually merge into one (fat) jet, as the two $h$ states can be very boosted.  %so that the actual appearance of such 
The final states in the detectors little resemble the primary parton kinematics of the underlying physics in such case. 
Finally, we assume the HL-LHC collider environment. This offers another challenge of increased presence of tracks in the final state not emerging from the aforementioned hard scattering and subsequent parton-to-jet dynamics, e.g., pile-up, soft underlying event, multi-parton scattering, etc.

The plan of this paper is as follows. In the next section, we describe the basics of a transformer encoder. Then, in Sect. 3, we introduce the physics process that we use as an example. Then, we present our numerical results. In Sect. 5, we interpret the classification results using various methods. The Sect. 6 is for conclusions.  The details of our network structure can be found in the appendix. %(We also have an Appendix describing the network structure.)

%%%%%%%%%%%%%%%%%%%%%%%%%%%%%%%%%%%
\section{Transformer encoder}
%%------------------------------_%%%
Transformers were originally proposed as sequence-to-sequence models for machine translation \cite{vaswani2017attention}.
%At that level, 
The main ingredient of the original transformer model is the encoder-decoder block. 
However, the models using encoder block only often appear for event classification analysis at the LHC. \cite{Kach:2022uzq,Finke:2023veq,Qu:2022mxj}. 

Inherited by the word tokens in the original transformer model,  transformer encoders are used to analyze events in terms of clouds of particles for High Energy Physics (HEP) analysis\cite{Komiske:2018cqr,Qu:2019gqs}. 
%Analogous to the point cloud representation, 
Particle clouds represent the final state particles as a permutation invariant sequence of particles. Such a representation has the ability to share %all 
the advantages of particle based representations, especially the flexibility to include arbitrary features for each particle. 

The motivation to apply transformer encoders to particle clouds stems from their inherent ability to model interactions between particles irrespective of their spatial proximity. By leveraging self-attention mechanisms, transformer encoders enable each particle to dynamically weigh the influence of other particles within the entire cloud, thus capturing both local and global dependencies. This can potentially revolutionize the analysis of HEP systems, particularly by offering a more holistic understanding of their behavior and interactions.

Understanding the scientific operation of transformer encoders in the context of particle clouds requires diving into the core components of these models. At the heart of the transformer architecture is the attention mechanism, an algorithm that allows the model to focus on different parts of the input sequence when making predictions.
An attention mechanism operates by assigning attention weights to different particles based on their relevance to the current particle being processed. This allows the model to consider global relationships and dependencies, enabling it to capture emergent behaviors, interactions, and patterns that may not be apparent %when using 
in filter based methods, e.g., Convolutional Neural Networks  (CNNs), which mainly extract the local information. 

%%%%%%%%%%%%%%%%%%%%%%%%%%%%%%%%%
\subsection{Attention mechanism}
%%%%%%%%%%%%%%%%%%%%%%%%%%%%%%%%%%%
The attention mechanism is an essential component of transformer models, playing a crucial role in capturing information and dependencies amongst particles. 
In the transformer architecture, the attention mechanism enables the model to focus selectively on different parts of the input sequence, allowing for the modelling of complex relationships and dependencies. 
In general, the attention mechanism operates by assigning different weights to different elements in the input sequence, emphasizing the more relevant parts while downplaying the less relevant ones\footnote{
Dropping the less informative instances from the data can rectify the sparsity problem when using CNNs to analyze jet images.}. The attention mechanism broadly span two types, as follows.
\begin{itemize}
    
    \item \textbf{Self-attention}  is a more advanced form of attention where the model attends to different positions in the input sequence to weight their importance concerning the current position. %Thus, 
    In the context of the transformer model, self-attention allows each element in the sequence to attend to all other elements, capturing both local and global dependencies. 
    Attention scores are calculated and used to combine the values associated with different positions linearly.
    
    The self-attention mechanism enables the model to consider the entire context, making it particularly effective for tasks where long-range dependencies are crucial.

    \item \textbf{Cross-attention} extends the self-attention mechanism to handle input sequences from different sources. In the transformer architecture, it is often used when processing pairs of sequences of different structures. Cross-attention allows each element in the first sequence to attend to all other elements in the subsequent sequence. This facilitates modeling the relationships between different modalities or extracting the relevant information from sequences with different scales.
\end{itemize}

Consider the  input data sets ($x_i,x_j$) that have first been passed by a linear fully connected NN layer to generate the weight matrices as follows: 
\begin{equation}
    Q_i = W^Q\cdot x_i \,,\hspace{4mm} K_j = W^K\cdot x_j \,, \hspace{4mm}  V_j = W^V\cdot x_j \,,
\end{equation}
where $K,Q$ and $V$ are called key, query, and value vectors, respectively, and used to compute the attention to the whole data set.

Scaled dot-product attention can then be defined as 
\begin{equation}
    \alpha_{ij} = \text{softmax}\left(\frac{Q_i\cdot K^T_j}{\sqrt{d}}\right) = \frac{\text{exp}(Q_i\cdot K^T_j/\sqrt{d})}{\sum_j\text{exp}(Q_i\cdot K^T_j/\sqrt{d})} \,,
    \label{eq:alpha}
\end{equation}
while the attention output is computed as a weighted sum of the attention scores as 
\begin{equation}
    \mathcal{Z}_i = \sum_j \alpha_{ij}V_j\,.
    \label{eq:attoutput}
\end{equation}
This is called self-attention if the attention is computed for the same data set, i.e., $x_i = x_j$.  The weights matrices have the dimensions of  $W_Q^{i\times i}, W_K^{i\times i}, W_V^{i\times i}$ which mixes the features of the input data and retain the dimension of the embedded input to the original one.   
In contrast, if the two input data sets differ, i.e., $x_i\neq x_j$,  cross attention is needed. 
In this case, the weight matrices should have different dimensions with $W_Q^{i\times i}, W_K^{j\times i}, W_V^{j\times i}$ in order to calculate attention.

Attention output is used to scale the input data set via a skip connection as 
\begin{equation}
    \widetilde{x_i} =  x_i + \mathcal{Z}_i \,.
    \label{eq:scaled_input}
\end{equation}
The new transformed data set  $\widetilde{x_i}$ indicates the attention importance of each element in the data set to the whole elements in the set. Although the attention output mixes the input and feature tokens, the skip connection keeps the reference to the order of the original input data set.

At its basic level, each transformer layer includes a multi-head attention, which combines different attention heads, allowing for parallel multi-dimensional processing of the inputs.
%In fact, 
Multi-head attention is a key innovation in the transformer model architecture, enhancing expressive power and capturing complex patterns in data by allowing the model to attend to different aspects of the input sequence simultaneously. 
Therefore, this mechanism eases the understanding of varied and subtle connections within the data, offering a more thorough representation.

As explained, a single set of attention weights is computed for the entire input sequence. 
Multi-head attention extends this concept by employing multiple attention heads, each responsible for learning different aspects of the relationships within the data. 
Each attention head independently processes the input sequence,  producing a set of output values. 
These outputs are then linearly combined to form the final output of the multi-head attention layer.

Mathematically, if $h$ represents the number of attention heads, and $\text{head}_i$ denotes the $
ith$ attention head, the output $\mathcal{O}$ is obtained by concatenating the outputs of each attention head and linearly transforming these:
\begin{equation}
    \mathcal{O} = \text{CONCAT}\left(\text{head}_1,\text{head}_2, \cdots \text{ head}_h \right) W^O\,,
\end{equation}
with $W^O$  the learnable linear transformation matrix which has the dimensions of $W_O^{ (h\ast i)\times i}$ to retain the same dimensions as the original input. 
This enables the model to capture different aspects of relationships and dependencies simultaneously.

The choice of the number of attention heads, denoted as $h_i$, is a crucial hyperparameter in designing a transformer model. 
Increasing the number of attention heads has several implications, such as enhancing the model's capacity to capture complex relationships.
It is also important to mention that a higher number of attention heads also increases computational complexity. 
Training and inference times and memory requirements could increase. 
Therefore, the number of attention heads should be balanced based on the task requirements and available computational resources.

\begin{figure}[h!]
\centering
 \includegraphics[width=10cm,height=8cm]{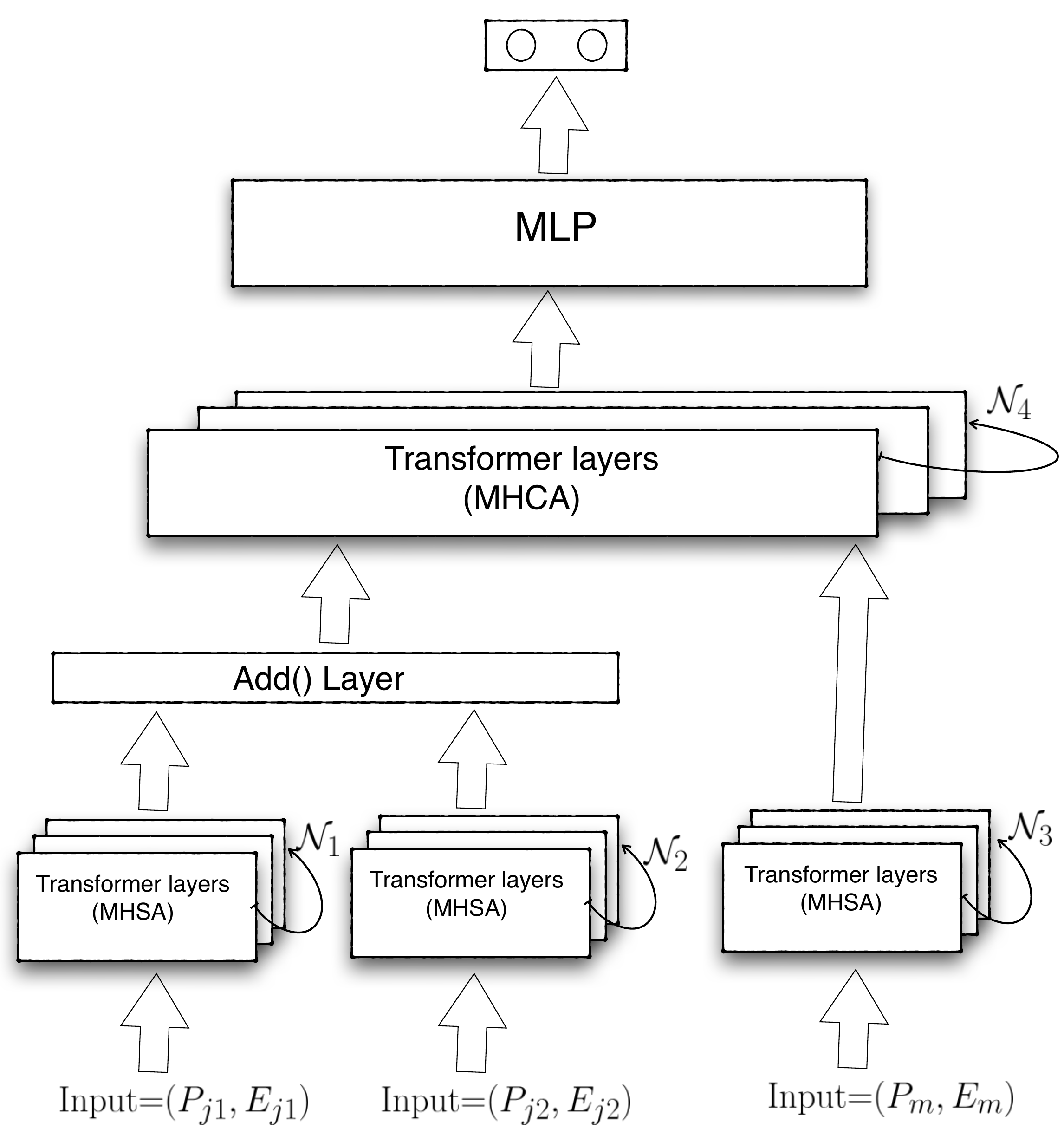}
    \caption{Structure of the transformer model used. Here, $P_{j1}, P_{j2}$ are the number of the leading and second leading jet constituents while the $P_m$'s are the reconstructed particles, $j_1,j_2$, and $H$. Also, 
    MHSA stands for multi-heads self-attention layers, and MHCA stands for multi-heads cross-attention layers. Finally, the  $\mathcal{N}_i$'s are the number of the used transformer encoders. The transformer layers are stacked and work sequentially, as pointed out by the black arrow.}
    \label{fig:network}
\end{figure}

In this particular context, we present an innovative methodology aimed at integrating inputs characterized by distinct scales within a multi-modal transformer model featuring cross-attention layers. The schematic representation of the network architecture is shown in Fig.~\ref{fig:network}. Considering the specific case of the HEP process to be studied at the LHC, $gg\to H \to hh\to b\bar b b\bar b$, the model dynamically adjusts three streams through self-attention transformer layers, each devoted to analyzing the leading jet, second-leading jet and the reconstructed kinematics, respectively.  At this juncture, the model independently extracts pivotal information from each data set, leveraging self-attention mechanisms before their collective processing through a cross-attention layer. 

 The main role of the cross-attention layer is to extract  the local jet substructure information effectively  and incorporate it into the extracted
 kinematic information. 
 Notably, the adaptability of the cross-attention layer in merging information from one data set into another affords flexibility in determining how to integrate the extracted information, providing the option to accentuate jet information for enhancing kinematics. 
 Once the most relevant information from the data sets is extracted and combined via the cross-attention layer, we feed the output to fully connected NNs to analyze the captured information and compute the classification probability.  
 The inclusion of self-attention layers in the model holds significance, as it allows for the independent extraction of the most relevant information from each data set prior to their amalgamation using the cross-attention mechanism. 
 This characteristic makes the model proficient in analyzing multi-scale data characterized by intricate structures.
 
%%%%%%%%%%%%%%%%%%%%%%%%%%%%%%
\section{Physics example}
\begin{figure}[th!]
\centering
 \includegraphics[scale=0.17]{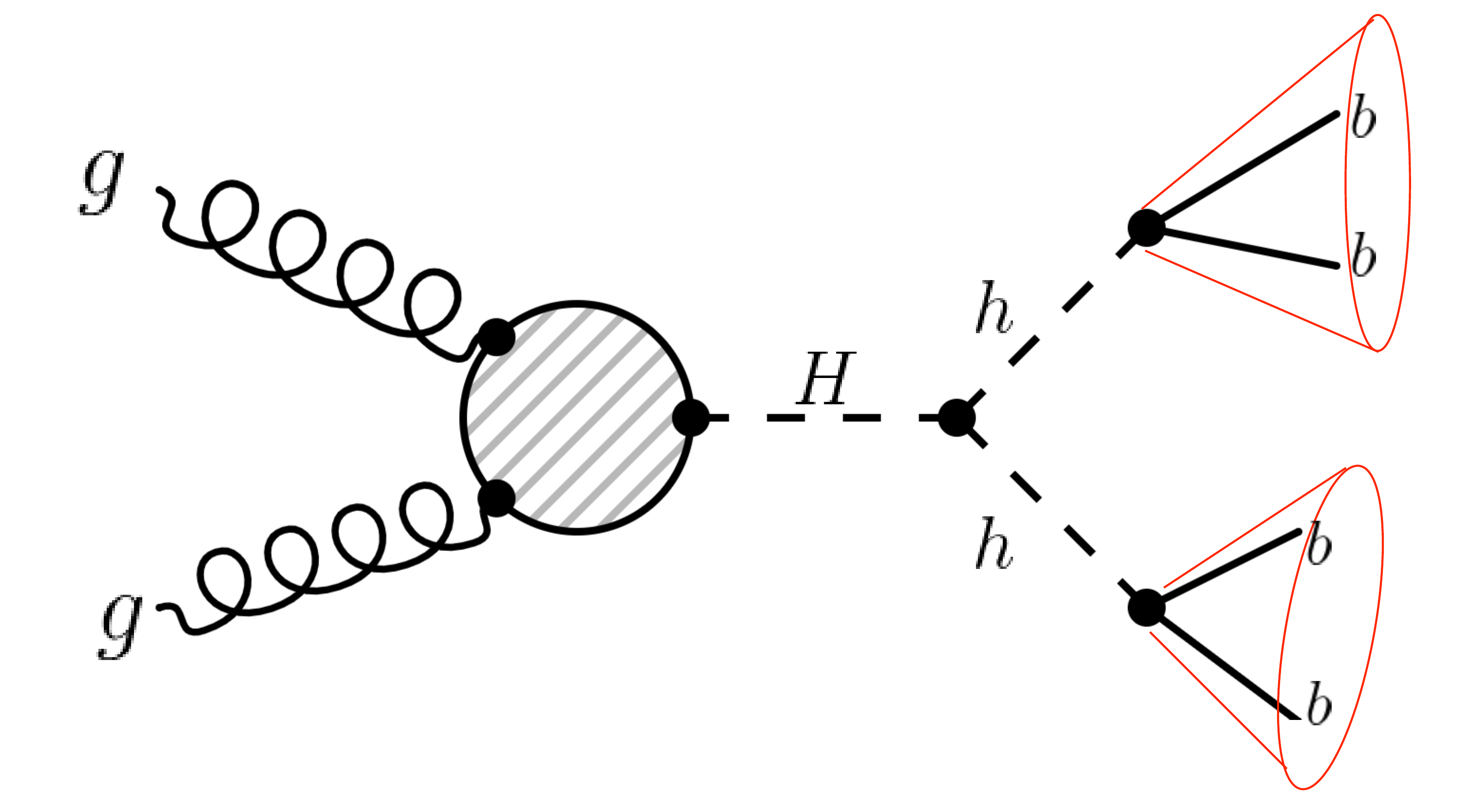}
    \caption{Feynman diagram for the signal process.}
    \label{fig:feyn}
\end{figure}

%%%%%%%%%%%%%%%%%%%%%%%%%%%%%
We analyse SM-like di-Higgs boson ($hh$) production at the HL-LHC with an integrated luminosity of $3000$ fb$^{-1}$ within the framework of the 2HDM. In the boosted regime, where the di-Higgs boson is produced from an on-shell heavy Higgs, $H$, the final state features two fat jets, as illustrated in Fig.~\ref{fig:feyn} by the two red cones therein.

Therefore, to start with, in this section, we provide a brief review of the 2HDM with type-II Yukawa couplings, focusing on the aspects that are relevant to our analysis. We then describe the strategy behind our numerical analysis, together with its constituent elements, i.e.,  the event generation and detector simulation procedures, as well as the signal and background properties, in terms of the overall kinematics and internal dynamics of jets. We adopt different transformer encoder configurations to analyze the kinematics and jet substructure individually and efficiently combine the information from both of these.

%%------------------------------
\subsection{The 2HDM}
%%------------------------------
The  2HDM is an extension of the SM  through a second $SU(2)_L$ Higgs doublet with the same quantum numbers under the SM symmetry gauge group \cite{Branco:2011iw,Lee:1973iz}. The most general 2HDM Higgs potential is given by
\begin{equation}
\begin{split}
    V_\phi &= m^2_{11} (\phi^\dagger_1 \phi_1) + m^2_{22} (\phi^\dagger_2 \phi_2) - \left[m^2_{12}(\phi^\dagger_1\phi_2)+\text{h.c.}\right]\\
     &+\lambda_1 (\phi^\dagger_1 \phi_1)^2+\lambda_2 (\phi^\dagger_2 \phi_2)^2  +\lambda_3 (\phi^\dagger_1 \phi_1) (\phi^\dagger_2 \phi_2) +\lambda_4 (\phi^\dagger_1 \phi_2) (\phi^\dagger_2 \phi_1)  \\
      &+\frac{1}{2}\left[ \lambda_5 (\phi^\dagger_1\phi_2)^2+ \left[\lambda_6(\phi^\dagger_1\phi_1)+ \lambda_7(\phi^\dagger_2\phi_2)\right](\phi^\dagger_1\phi_2)+\text{H.c.}\right]  \,.
\end{split}
\label{eq:Vphi}
\end{equation}

The potential structure allows for Flavor Changing Neutral Currents (FCNCs) at the tree level,  which is strongly constrained by experimental measurements. Applying a global $Z_2$ symmetry to the scalar potential, with $(\phi_1,\phi_2)\to(\phi_1,-\phi_2)$ transformations, prevents the existence of such FCNC sources \cite{Glashow:1976nt}. 
However, the most general Yukawa interaction violates such a $Z_2$ symmetry, thus leading to potential FCNCs at the tree level, as pointed out in Ref. \cite{Ginzburg:2004vp}. 
{Therefore, to tame the latter, only specific Yukawa structures, known as Types \cite{Branco:2011iw}, are allowed. Yet, to enable 
Electro-Weak Symmetry Breaking (EWSB) consistent with the measured particle spectrum of the SM, a softly broken $Z_2$ symmetry should eventually be enabled by requiring a small but non-vanishing term  $m^2_{12}(\phi^\dagger_1\phi_2)$ and setting $\lambda_6=\lambda_7=0$. (Herein, softly means that the model still respects the $Z_2$ symmetry at small distances through all orders of perturbation theory.)  The soft mass $m^2_{12}$ and $\lambda_5$ are in general complex, though \cite{Antusch:2020ngh,Antusch:2021oit}. In the following, we will consider a real potential that thus preserves the CP symmetry, ${\rm IM}(m^2_{12})={\rm Im}(\lambda_5)=0$. In such a configuration of the 2HDM, then 7 independent parameters remain, which are the $\lambda_i$'s, with $i=1,\dots 5$, $\tan\beta=v_2/v_1$\footnote{With $v_1$ and $v_2$ being the Vacuum Expectation Values (VEVs) of the two Higgs doublets.} and $m_{12}^2$, from which the physical parameters, i.e., Higgs boson masses and couplings, are obtained, with the constraint that one of the two CP-even Higgs fields should be the discovered one with mass of 125 GeV or so (which in our case is the $h$ field). Finally, as mentioned already, we restrict our study to the Type-II among the possible Yukawa structures.
}

The tree level mass matrix squared for the Higgs fields can be obtained as
\begin{equation}
    \left(\mathcal{M}^2 \right)_{ij}= \frac{\partial V_\phi}{\partial h_i\partial h_j} \Bigg|_{h_{i,j}=0}\,,
\end{equation}
where the $h_i$'s ($i=1,\ldots ,4$) are the four components of the complex doublet fields. Upon EWSB, three physical neutral scalars are obtained after diagonalizing the corresponding mass matrices, as intimated, two CP-even (scalar) ones ($h, H$) and a CP-odd (pseudoscalar) one ($A$),  with masses given by
\begin{align}
    \label{eq:massh1h2}
    m^2_{h,H} & = \frac{1}{2}\left[\chi^2_{11}+\chi^2_{22}\mp \sqrt{(\chi^2_{11}-\chi^2_{22})^2+4(\chi^2_{12})^2}   \right]\,, \\
    \label{eq:masshA}
    m^2_A  &= \frac{2m^2_{12}}{\sin2\beta}-\lambda_5 v^2\,,
\end{align}
with
\begin{align}
    \label{eq:mass11}
    \chi^2_{11} & =m^2_{12}\tan\beta+2\lambda_1 v^2\cos^2\beta\,,\\
    \label{eq:mass22}
    \chi^2_{22} & =m^2_{12}\cot\beta+2\lambda_2 v^2\sin^2\beta\,,\\
    \label{eq:mass12}
    \chi^2_{12} & =-m^2_{12}+\frac{1}{2}(\lambda_3+\lambda_4+\lambda_5) v^2\sin2\beta\,,
\end{align}
where the VEVs satisfy the relation $v=\sqrt{v_1+v_2}$ (with $v$ being the SM one)\footnote{The other two Higgs states emerging from the 2HDM after EWSB are charged and are denoted by $H^\pm$.}.

To stay with the neutral Higgs sector, the imposed CP conservation only allows for tree level couplings between two massive gauge bosons and the CP-even Higgs states, while the CP-odd Higgs state can only couple to a gauge boson and a CP-even Higgs one. Furthermore, all neutral Higgs states can couple to fermions. The coupling strength of the neutral Higgs bosons to both matter and forces are parameterized in terms of $\tan\beta$ and another parameter, $\alpha$, which is the mixing angle between the CP-even Higgs states \cite{Branco:2011iw}. Furthermore, the triple scalar coupling is independent of the Yukawa structure and is given by \cite{Arhrib:2009hc}
\begin{equation}
    \lambda_{(H,h,h)} = -\frac{e\ c_{\beta-\alpha}}{2m_W s_W s^2_{2\beta}}\left[(2m^2_h+m^2_H) s_{2\alpha}s_{2\beta}+(3s_{2\alpha}-s_{2\beta})m^2_{12}\right]\,,
\end{equation}
where $e$ is the electric charge and $s,c$ are the $\sin$ and $\cos$ of the given angle.

The 2HDM free parameters are constrained from various theoretical considerations and
experimental observations, as described in \cite{Hammad:2022wpq}. Thus, profiting from the scan results performed therein, we adopt four Benchmark Points (BPs), with $m_H= 600, 800, 1000$, and $2000$ GeV, that satisfy all the current bounds. 
In Tab.~\ref{tab:tab_1}, we show the parameters values of these points while the last column shows the production cross section $\sigma_{\rm prod}$ of our target process (prior to the two $h\to b\bar b$ decays) 
at $\sqrt{s}= 14$ TeV. 

\begin{table}[h!]
\caption{Input parameters for our four BPs. The last column shows the production cross section for the process $gg\to H\to hh$.}\vspace*{0.25cm}
\begin{tabular}{|p{0.085\textwidth}|p{0.04\textwidth}|p{0.04\textwidth}|p{0.04\textwidth}|p{0.06\textwidth}|p{0.06\textwidth}|p{0.06\textwidth}|p{0.11\textwidth}|p{0.12\textwidth}|p{0.1\textwidth}|}
\hline
$m_H$[GeV] & $\lambda_1$ & $\lambda_2$ & $\lambda_3$ & $\lambda_4$ & $\lambda_5$ & $\tan\beta$& $m^2_{12}$[TeV$^2$] &$\cos(\beta-\alpha)$  & $\sigma_{\text{prod}}$[fb] \\ \hline
$600$ & $1.80$&$0.23$&$1.75$&$-2.06$&$-1.09$&$5.00$&$-78.97$&$0.31$& $0.86$ \\  \hline
$800$ & $3.20$ & $0.25$ & $1.75$ & $-2.06$ &$-1.29$& $4.00$ &$-128.91 $ &$0.33$ &$0.375$  \\  \hline
$1000$ & $1.0$ & $0.16$ & $3.50$ & $ -2.06$ &$-1.49$& $3.00$ &$-302.92$ &$0.37$ &$0.11$  \\  \hline
$2000$ & $1.0$ & $0.14$ & $2.75$ & $-1.06$ &$-1.97$& $5.00$ &$-889.05$ &$0.32$ &$0.024$  \\  \hline
\end{tabular}
\label{tab:tab_1}
\end{table}

%%------------------------------
\subsection{Analysis strategy}
%%------------------------------
With the theoretical setup clarified, we now proceed to a phenomenological study of di-Higgs boson production, focusing on final states with two boosted fat jets. 
%In this approach, 
We align our analysis with the boosted analysis presented in the latest ATLAS paper \cite{ATLAS:2022hwc}. 
The primary background contamination arises from QCD multijet processes, specifically $pp\to jjjj$, contributing an estimated $90\%$ of the total background, while the di-top process $\bar t t$ contributes at the  $10\%$ level. 
Other background processes, including SM $h$, $hh$, and EW di-boson production, have been assessed to make negligible contributions to the selected event yields. 
Therefore, they are not included in our analysis. 
Given that BSM di-Hggs events suffer from huge background contamination and it is not trivial to extract the signal information, we employ various configurations of transformer encoders for this analysis.

Commencing with the analysis of the global information encoded in the high-level reconstructed kinematics of both the signal and relevant background events, we employ a transformer encoder with multi-head self-attention to optimize the separation power between signal and background events. 
However, the presence of similar (to the signal) kinematic structures in some background processes poses a challenge to the classification efficiency of this network. 
Additionally, the substantial cross section of the background events diminishes signal significance, even after optimizing the cut on the output score.
To enhance the network performance, one may consider applying initial cuts on certain variables before inputting the distributions into the network, aiming to amplify the signal and suppress the backgrounds. 

In this analysis, we adhere to the pre-selection cuts outlined in \cite{ATLAS:2022hwc}, requiring two fat jets with a double $b$-tagging each. 
Moreover, each event must have at least two fat jets with radius $R=1.0$ and $pT> 450$ GeV for the leading jet and $pT > 250$ GeV for the second leading jet. 
Each of the two fat jets is required to have a pseudorapidity $|\eta(J)| < 2.5$, a lower mass cut of $m(J) > 50$ GeV, and a mass window of $200$ GeV is applied for the $m_H$ reconstruction for $m_H \le 1$ TeV and relaxed for higher masses to allow for more statistics.  
Unlike the ATLAS analysis, we do not consider pile-up effects in this analysis.   

In addition to the global kinematical variables, we can utilize jet substructure %methods 
of the jets to distinguish between signal and background events. This naturally arises from the fact that jets initiated by different particles exhibit distinct characteristics. 
While heavy boosted particles, such as $W^\pm$, $Z$ and Higgs bosons, can result in jets with a distinctive multi-prong structure, quark and gluon jets are unlikely to have such structure. 
Furthermore, the boosted colour singlet particle is isolated in colour flow. Therefore, two b jets from Higgs decay are colour-connected only among themselves, unlike QCD jets. 

Consequently, the features of the parent particles can be inferred from the structure of the jet constituents. This information enables the recovery of various local details about events from different processes, serving as a discriminating variable between signal and background events. The study of jet substructure to identify the parent particle initiating a jet, thereby  
Distinguishing jets initiated from heavy boosted particles from QCD jets was introduced in \cite{Kaplan:2008ie,Cui:2010km,Plehn:2011sj,Soper:2012pb,Anders:2013oga,Kasieczka:2015jma,Thaler:2010tr,Thaler:2011gf,Larkoski:2013eya,Moult:2016cvt,Larkoski:2014wba,Abdesselam:2010pt,Altheimer:2012mn,Altheimer:2013yza} (and  references herein). Recently,  improvement on jet identification continued by using ML methods for jet image analysis \cite{Cogan:2014oua,Almeida:2015jua,deOliveira:2015xxd,Baldi:2016fql,Barnard:2016qma,Komiske:2016rsd,Kasieczka:2017nvn,Macaluso:2018tck,Choi:2018dag}, graph based analysis \cite{Mokhtar:2022pwm,Ma:2022bvt,Gong:2022lye} or sequence based analysis \cite{Guest:2016iqz,Pearkes:2017hku,Egan:2017ojy,Fraser:2018ieu,Butter:2017cot,Kasieczka:2018lwf}. Exploiting the jet substructures of the Higgs jets to discriminate between signal and background events, 
In this paper, we especially employ a multi-modal transformer encoder with self-attention multi-heads to analyze the jet contents. The different modalities are designed to extract information from the leading and second-leading jet contents in parallel before a simple concatenation is performed for classification purposes. 
Without cross attention to the high-level kinematical variable discussed next, the classification performance is based solely on the information localized inside the fat jet.

Integrating inputs of varying scales encompassing both kinematics and jet substructure information, we utilize a multi-modal transformer encoder equipped with three streams and cross-attention head. 

The first and second streams process information from the leading and second-leading jet contents. Each of them features a transformer encoder with self-attention heads. 
Once important features are extracted from the jets, they are aggregated in an addition layer. The third stream, dedicated to high-level kinematics, employs a transformer encoder with self-attention heads. The output from the addition layer and the final layer of the third transformer are fed into a cross-attention layer. 
This cross-attention layer is pivotal in connecting information extracted from the jet constituents to the corresponding kinematics, enhancing the overall classification performance. To elucidate the impact of the cross-attention layer, we introduce a fourth model wherein we substitute the cross-attention layer with a straightforward concatenation layer.

For events simulation, we use MadGraph5 \cite{Alwall:2014hca} to estimate multi-parton amplitudes and to generate signal and background events for subsequent processing. 
Background processes $pp\rightarrow bbbb$ \footnote{We simulate the multi-jets QCD processes in Madgraph as $pp\rightarrow b\bar{b} b\bar{b}$ to speed up the event generation.} and $pp\rightarrow t\bar{t}$ are computed at Leading Order (LO)  while the Higgs production from gluon-gluon fusion is calculated at Next-to-LO (NLO) in QCD using an effective coupling calculated by SPheno \cite{Porod:2003um,Porod:2011nf}. PYTHIA \cite{Sjostrand:2006za} is used for parton shower, hadronization, heavy flavor decays, and for adding the soft underlying event. 
The DELPHES package \cite{deFavereau:2013fsa} is used for detector simulation.

DELPHES parameterizes the detector response, including tracks, calorimeter deposits, and high level objects such as isolated electrons, jets, taus, and Missing $E_T$ (MET). We use the default ATLAS card for resolution, but the (fat)jets are reconstructed from the Eflow objects, combining tracks and calorimeter information. 
The $\bar{t}t$ background is simulated at LO and up to two more jets with the matching scale $20$ GeV via the MLM method \cite{Alwall:2007fs,Mangano:2006rw}. 
Fat jets are clustered using the anti-$k_T$ algorithm \cite{Cacciari:2008gp,Catani:1993hr}  with cone radius $R=1$ and, to ensure further suppression of pile-up noise,  jet trimming is performed \cite{Krohn:2009th}. 
%%------------------------------
\subsection{Data pre-processing}
\begin{figure}[th!]
    \includegraphics[scale=0.165]{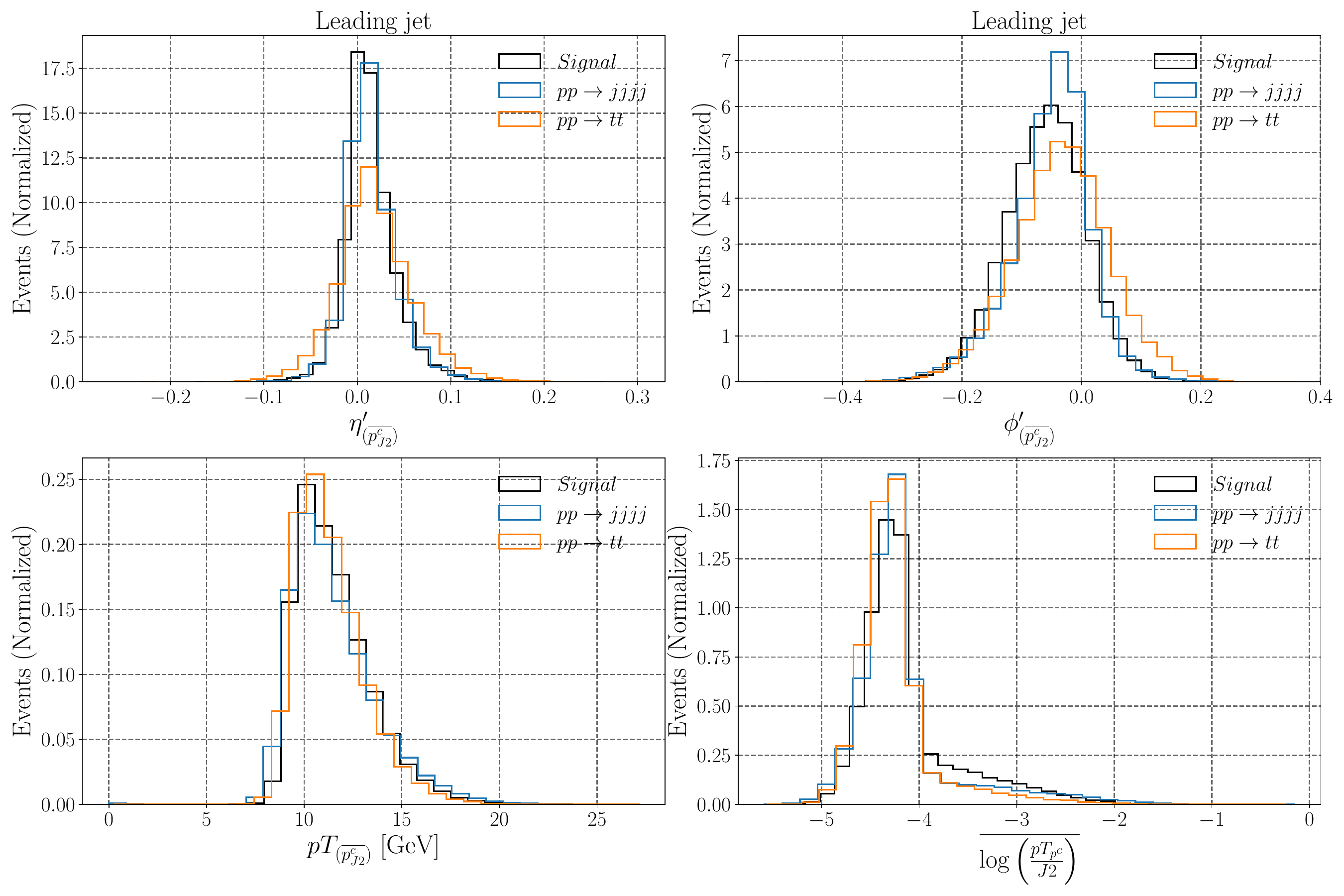}\includegraphics[scale=0.165]{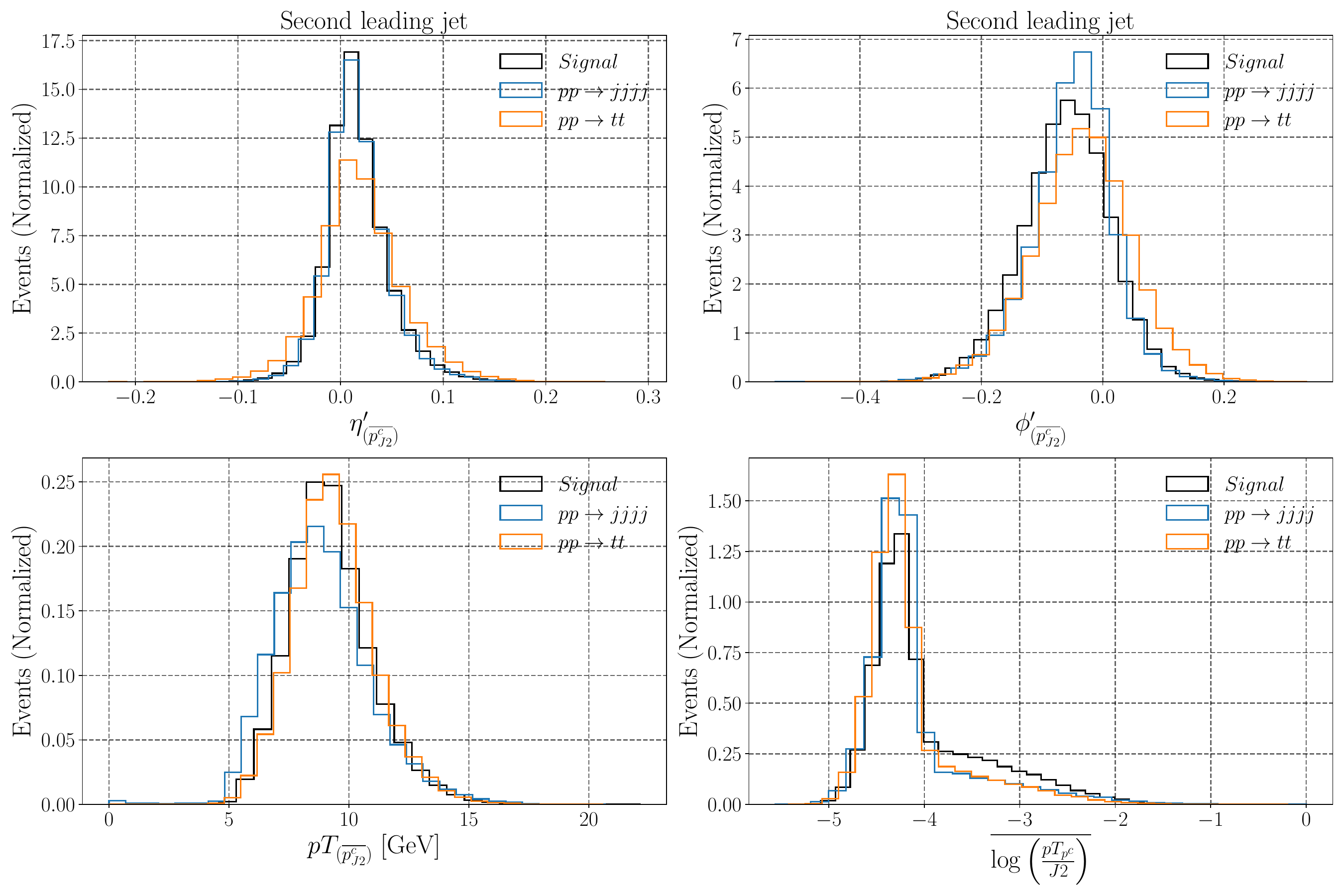}
    \caption{Left: distributions for the averaged constituents of $10000$ leading jets. Right: distributions for the averaged constituents of $10000$ second leading jets. Signal distributions are %considered 
    for the  BP with $m_H = 1$ TeV.   }
    \label{fig:jet_kin}
\end{figure}

%%------------------------------
Particle clouds enable configuring diverse data into the network, emphasizing the permutation symmetry of inputs to yield a promising representation of jets.
Initially, we pre-process the data sets for the leading and second-leading jet contents up to  $50$ constituents each. 
The particles are arranged in descending order based on their transverse momentum. 
For events with fewer constituents, the remaining positions are padded with zeros, ensuring conformity with the stipulated count\footnote{We stress here that we use an attention mask such that the network performance is not affected by the padded events\cite{vaswani2017attention}.}. For each instance of the jet contents we store $4$ features:  $pT,\eta,\phi$ and $\log\frac{pT}{pT_{(\text{jet})}}$\cite{Qu:2019gqs}. Fig. \ref{fig:jet_kin} shows the four features averaged over the number of jet contents for $10000$ events of the leading jet (left) and second leading jet (right).

To optimize the network discriminative accuracy, it is imperative to pre-process the jet contents, ensuring the manifestation of a multi-prong structure specific to signal events.
We use the preprocessing steps that were introduced for jet image analysis.
The preprocessing allows learning from small input data and considerably speeds up the learning process
\footnote{In principle, we can use the input data without the preprocessing, but this needs a large input data set and train for a long time \cite{Qu:2022mxj}}. For this purpose, the following transformations are applied before inputting the data into the network\footnote{Other than the rotation and flipping as proposed below, it is also possible to recluster the fat jet into subjets and define the rotation and flipping based on the subjet locations.}.

\begin{itemize}
    \item \textbf{Translation} Jet contents are shifted in the $\eta-\phi$ directions such that the jet axis is at the center of the $\eta-\phi$ plane.
 \item \textbf{Rotation} Rotation is executed to mitigate the stochastic nature inherent in the decay angle concerning the $(\eta-\phi)$ coordinate system. This alignment is achieved comprehensively by ascertaining the principal axis of the original data and subsequently rotation around the jet-energy centroid. This rotation ensures that the principal axis consistently aligns vertically. The rotation transformation is performed by first computing the leading eigenvector of the covariance matrix as the principle axis of the jet. A rotating angle, $\theta$,  is then defined as $\mathrm{arctan2}\left( \frac{x_1}{x_2}\right)$, with $x_1,x_2$ are the first and second components of the eigenvector respectively. Finally, the rotating angle is used to rotate the $(\eta-\phi)$ coordinates of the jet constituents to new non-physical coordinates,  $(\eta^\prime-\phi^\prime)$, in which the principle axis of the jet is always vertical.
 
\item \textbf{Flipping} Jet constituents are reflected over the vertical axis such
that the right side of $\eta^\prime$ always has the highest momentum.  This ensures that the hardest radiation always appears in similar locations, which can be exploited to enhance the classification performance.
\end{itemize}

After pre-processing transformations, input data sets for the leading and second leading jets have the dimensions of $(n,50,4)$, where $n$ is the number of events, $50$ is the number of jet constituents, and $4$ is the number of pre-processed features. 

%%%%%%%%
\begin{figure}[h!]
    \includegraphics[scale=0.25]{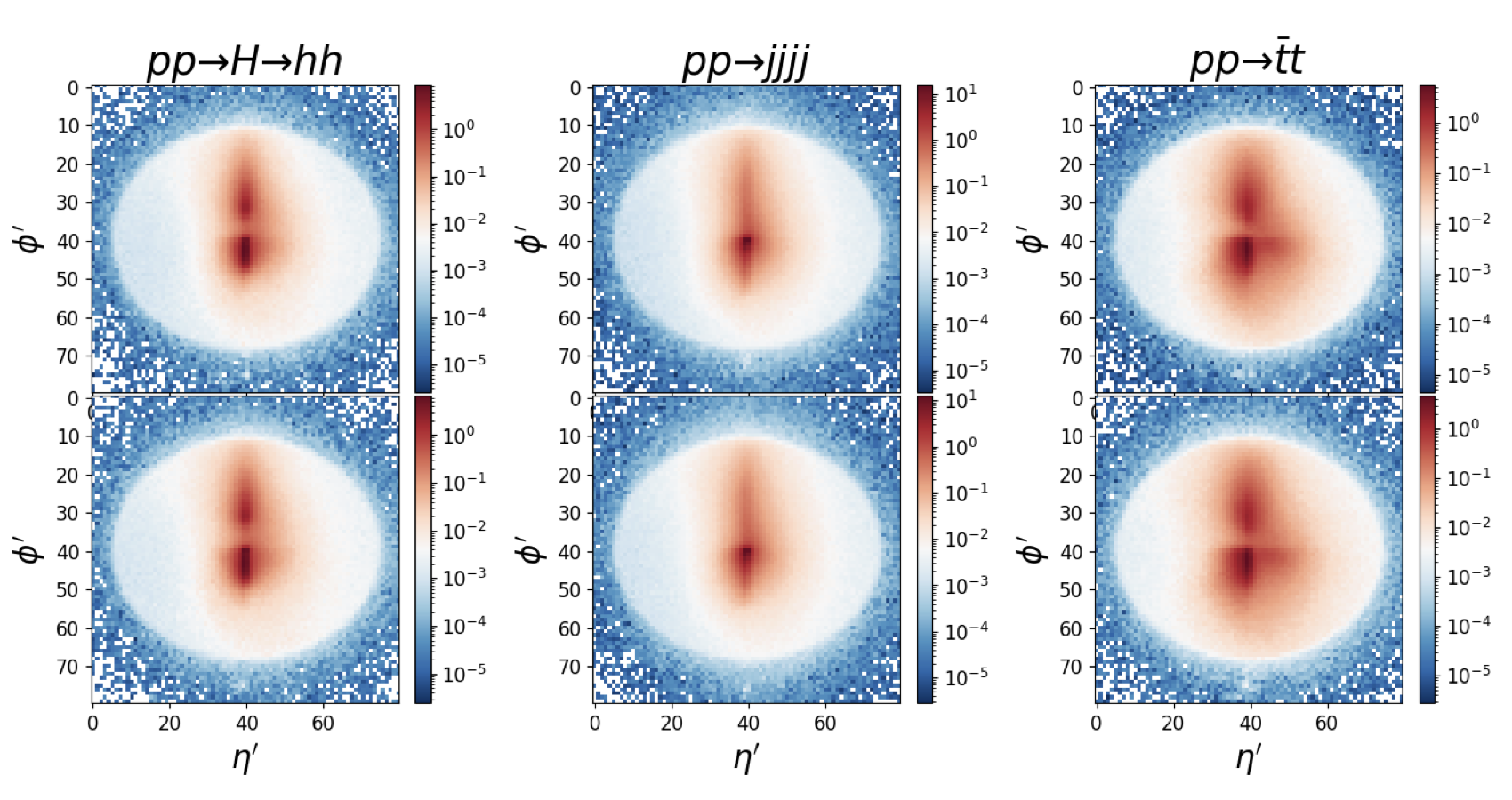}
    \caption{For illustration purposes, we show the accumulated average of $50000$ $p_T$ distributions of the leading (second leading) fat jet contents in the upper raw (lower raw)  after pre-processing steps for both signal and backgrounds. The signal events (left) are simulated for the BP with $m_H=1$ TeV and shown against the yield of the $bbbb$ (center) and $\bar t t$ (right) background events. Here, $X$ and $Y$ ticks indicate the bin in $\eta$ and $\phi$ direction.}
    \label{fig:jet_images}
\end{figure}

Fig.~\ref{fig:jet_images} presents the cumulative average of $50000$ $p_T$ distributions for both the leading (upper row) and second-leading (lower row) jets. The impact of pre-processing transformations is evident in revealing the multi-prong structure characteristic of signal events, wherein the leading and second-leading subjets are localized in specific regions within the $(\eta^\prime-\phi^\prime)$ plane. In contrast, subjets from QCD multi-jets exhibit a broad energy range, lacking a discernible prong structure. Conversely, $\bar{t}t$ events show a distinct three-prong structure attributable to the fully hadronic decays of the top quark. Notably, despite the multi-prong structure in $\bar{t}t$ background events, their contribution to the overall background is merely $10\%$. We will see later that background rejection efficiency is high, therefore $t\bar{t}$
background can be important to estimate the accessibility of the signal. 
%exerting minimal impact on the classification performance.

\begin{figure}[h!]
    \includegraphics[scale=0.29]{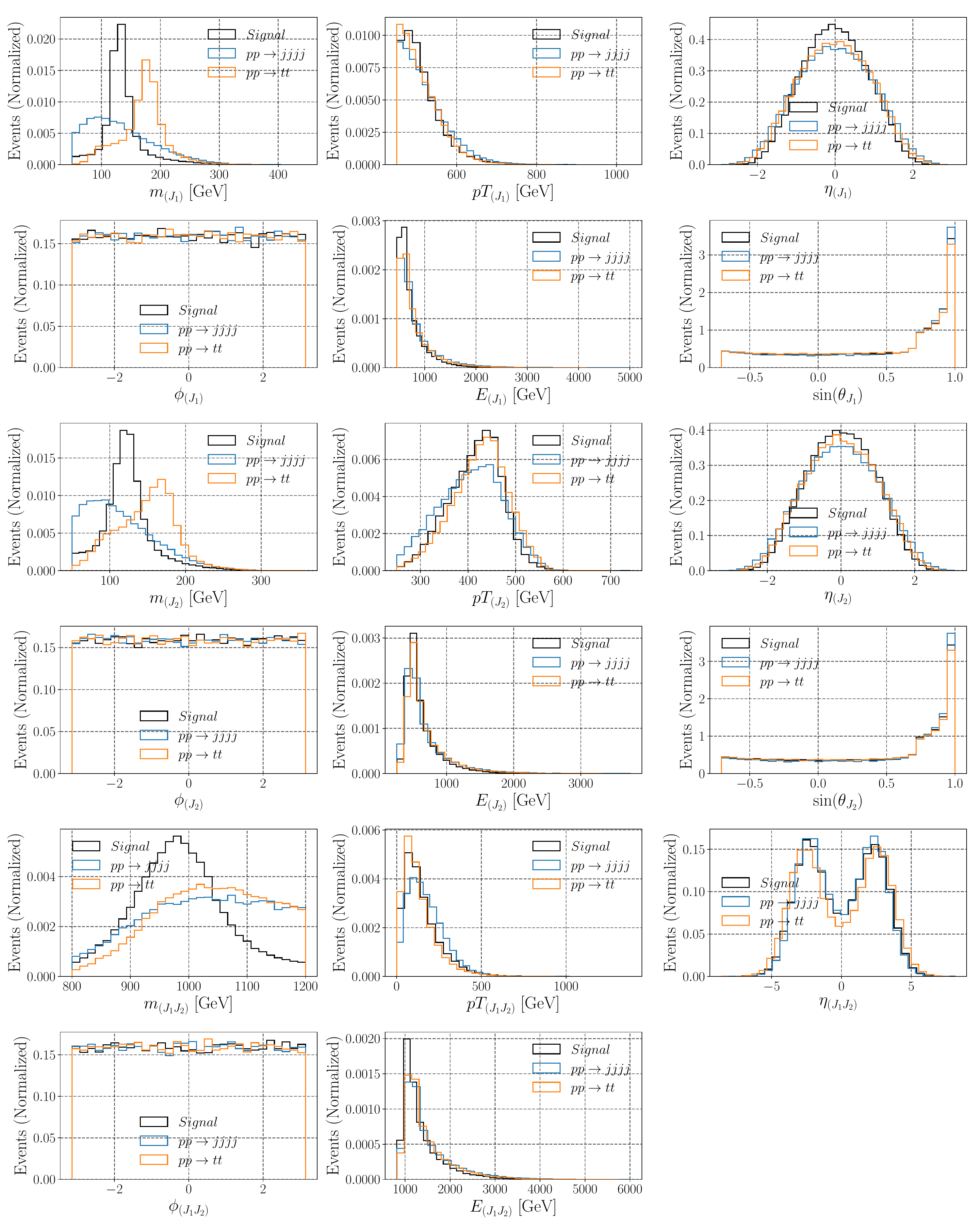} 
    \caption{Kinematics distributions of 10000 events for the signal BP with $m_H$ =1 TeV and the corresponding backgrounds after applying the pre-selection cuts.}
    \label{fig:kinematics}
\end{figure}

The kinematics data sets have dimension $(n,3,6)$ with $n$ as the number of events, $3$ as the number of reconstructed particles, leading, second leading jet, and heavy Higgs, and $6$ as the number of the kinematic variables for each reconstructed particle.  
The 6 kinematic variables are  mass $m$, $p_T$, $\eta$, $\phi$, and energy of the jet $E$ and the rotation angle of the jet $\theta$. 
Note that we assign 5 inputs corresponding to the 4 momenta of the jet. 
Because of the kinematical constraints $p^2=m^2$ and $p_H=p_{J1}+p_{J2}$, there are only 8 physically independent observable among 15 kinematical inputs. These additional inputs help the network to figure out relevant features for the classification. 

Fig. \ref{fig:kinematics} shows the normalized kinematic distributions for the signal point with $m_H=1$ TeV and backgrounds. In addition to the reconstructed high-level kinematics, we incorporate the $\theta_i$ distributions for the leading and second-leading jets (but not the heavy Higgs), which are the rotating angles of the leading and second leading jet contents. 

We incorporate the data sets as input to the networks as follows, inputs to the first and second transformer encoders have the dimensions of $(n,50,4)$. In contrast, the input to the third transformer encoder has the dimension of $(n,3,6)$. Once the data sets are pre-processed, we stack signal and background events in each data set separately, attaching labels of $Y=1$ for the signal events and $Y=0$ for the background events. %Once the labels are adjusted, 
During the training of the network, the model tries to minimize a categorical cross entropy loss function by minimizing the difference between the model prediction and the assigned labels. In this analysis, we use equal size data sets for signal and background events for training with $1 $ million events\footnote{A major problem in any attention based transformer model which exhibits larger classification performance with larger size training set.} and $100000$  event for test. 
%%%%%%%%--------
\section{Results}
%-----------------
We now present the analysis results for probing the signature of the heavy scalar in the process of boosted di-Higgs boson production, $gg\to H\to hh\to b\bar b b\bar b$, at the HL-LHC with integrated luminosity of $3000$ fb$^{-1}$. The discriminating power of each network will measure how well the signal and background may be characterised through their different features, all entangled together into several kinematic distributions and jet substructure information. For this purpose, we utilize four different attention based transformer models which analyze the reconstructed high level kinematics or the jet substructures individually via transformer encoders with self-attention mechanism. Alternatively, we adopt two multi-modals transformer encoders to analyze the combined information of kinematics and jets substructure. In the latter, we incorporate the different information using a simple concatenation layer or cross-attention layer. A full description of the used networks is in Appendix \ref{appendix:1}. 

\begin{figure}[th!]
    \includegraphics[scale=0.236]{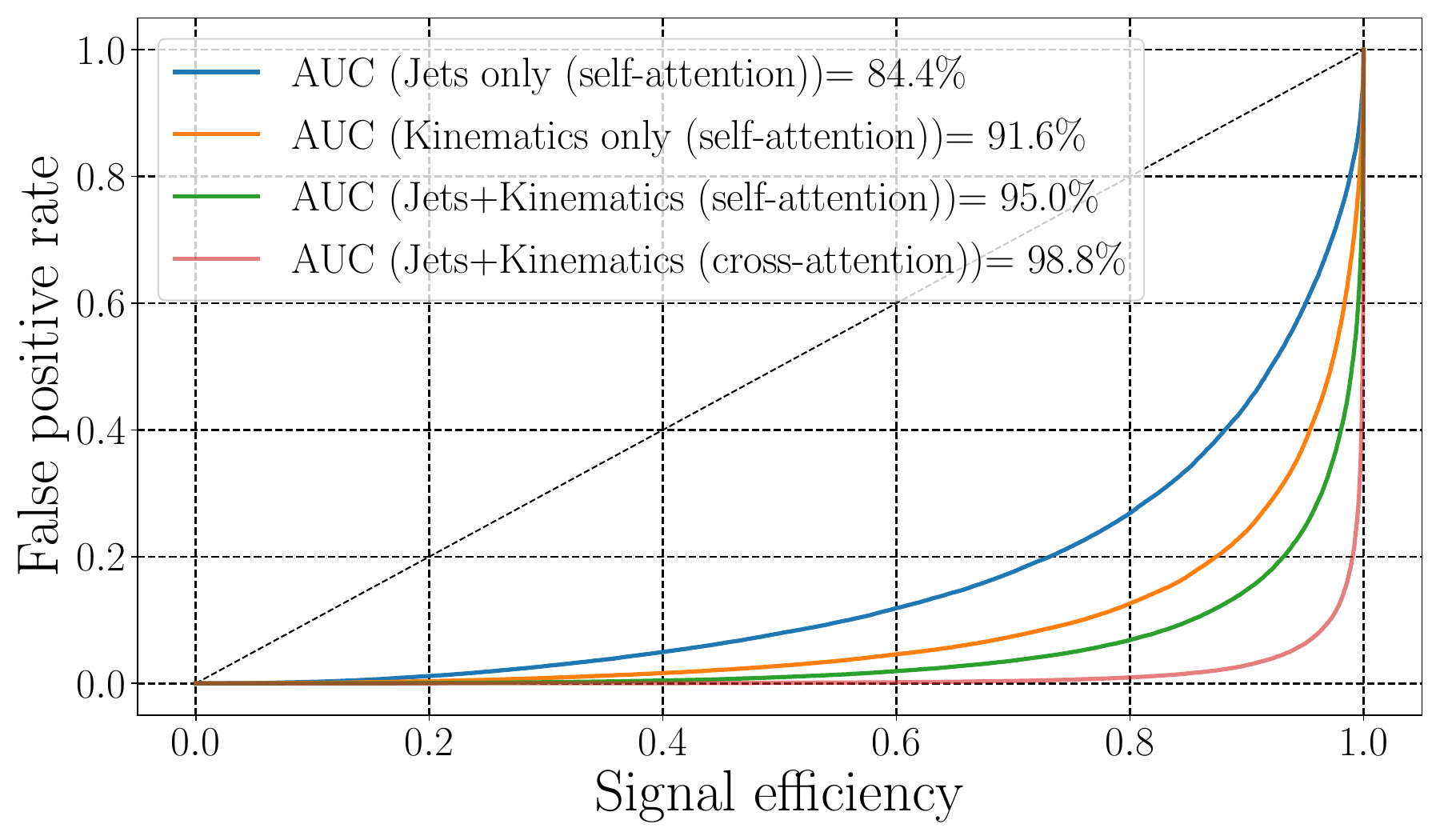}~\includegraphics[scale=0.25]{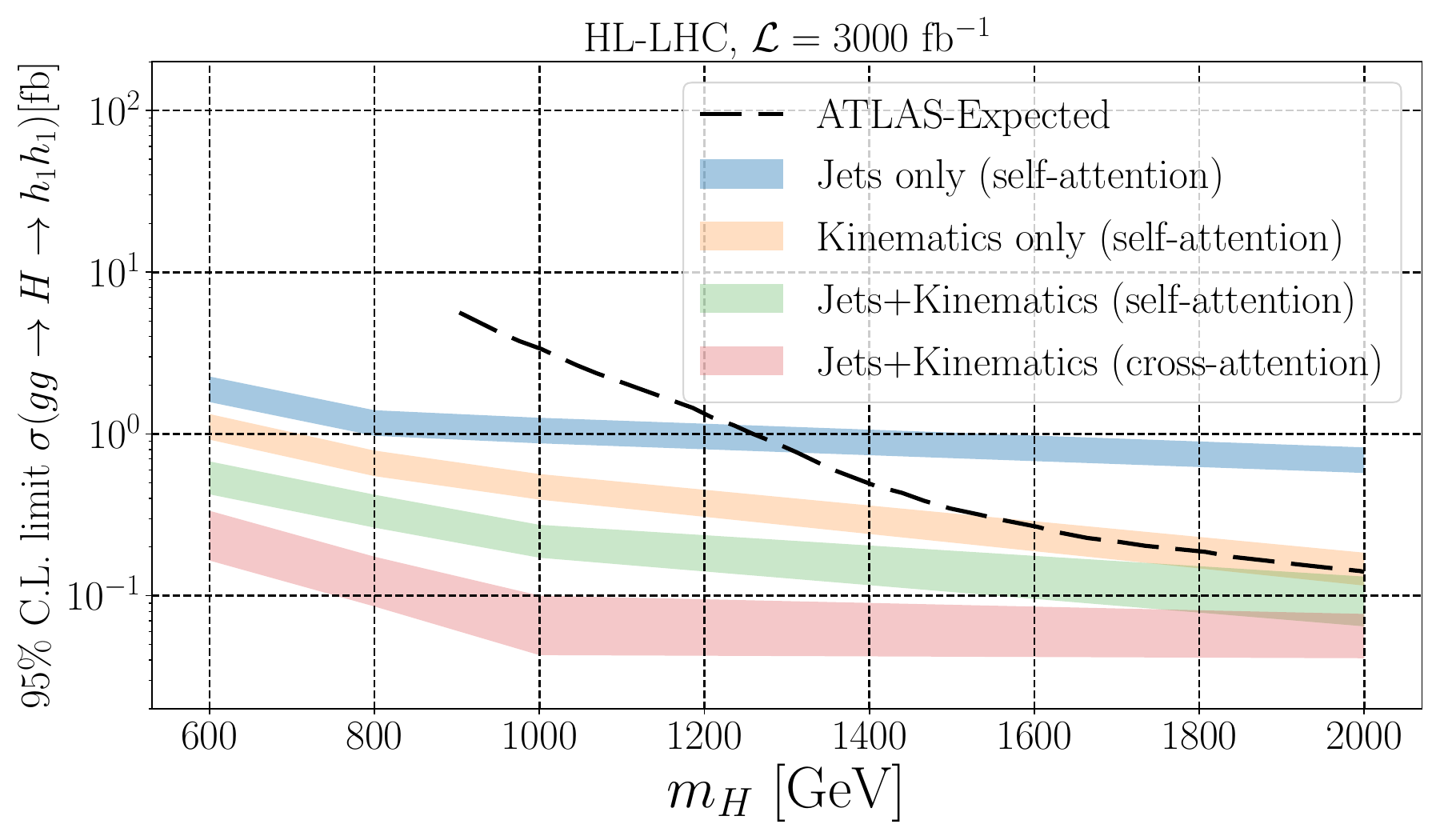}
    \caption{Left: The  Receiver Operating Characteristic
    (ROC) curves for the four networks for the signal BP with $m_H=1$ TeV. Right: $95\% $ upper limit on the total cross section for the process $gg\to H \to h h$  (having factored out the SM-like $h\to b\bar b$ decays) at the HL-LHC with integrated Luminosity $3000$ fb$^{-1}$ for different ML analyses.
    The band for each plot represents the upper and lower values for 5 independent training of different randum number seeds, and the middle line represents the central values. 
    The ATLAS limits are extracted from the latest analysis in \cite{ATLAS:2022hwc} and linearly scaled to the integrated luminosity of $3000$ fb$^{-1}$. }
    \label{fig:results}
\end{figure}

The classification performance of the utilized networks is presented in Fig.  \ref{fig:results}. In the left plot, we showcase the ROC for the employed networks for a signal with $m_H=1$ TeV. The multi-modal transformer encoder with cross-attention layers performs best, achieving an Area Under the Curve (AUC) of $98.8\%$. In contrast, the transformer encoder trained solely on the jet substructure information exhibits the lowest performance with an AUC of $84.4\%$. It is crucial to highlight the impact of the cross-attention layer, which enhances performance by $7\%$ over the transformer model trained exclusively on kinematic information. Replacing the cross-attention layer with a simple concatenation layer results in a degradation of classification performance by approximately $\sim 4\%$, as depicted by the green line in the plot.

In the right plot, we present the $95\%$ upper limit on the production cross-section at the HL-LHC for heavy scalar mass ranges between $600-2000$ GeV. The dashed black line represents the limit for the ATLAS analysis \cite{ATLAS:2022hwc}, with linear scaling of the integrated luminosity to $3000$ fb$^{-1}$. For lower masses, $m_H\le 1$ TeV, all the used transformer models show enhanced performance over the ATLAS analysis, exhibiting over $10$ times better sensitivity. For larger masses, for which the reconstructed kinematics of the signal are faithful to its true structure with vanishing background events, the performance of the transformer models saturates. In fact, for the limit, e.g., $m_H=2$ TeV, the background events can be easily removed with a simple cut on the reconstructed distributions of the signal events, which exhibits a clear difference from the background distributions. The transformer network trained on the jet constituents only does not show a large impact with varying the heavy scalar mass. 

The network performance is subject to training and statistical uncertainty from limited training and testing samples.  
For example,  the network performance can be influenced by the random partitioning of the training and test data sets, and the network performance varies when repeating the training and test steps with new splits. 
 
We repeat the experiment for $k$ times and report the results as bands between the highest and lowest values. In our results, we use $k=5$, and the bands represent the values of the different represented experiments.

As for optimizing the signal-to-background yield, we enforce a cut on the network output score to keep only $20$ events of the background. With this choice, we alleviate the statistical errors that may occur for lower background\cite{Cowan:2010js}. 

The optimized signal and background events are used to derive the upper limit using the following formula \cite{LHCDarkMatterWorkingGroup:2018ufk}
\begin{equation}
Z_A = \left[ 2\left( (N_s+N_b)\ln\frac{(N_s+N_b)(N_b+\sigma^2_b)}{N_b^2+(N_s+N_b)\sigma^2_b}  -\frac{N^2_b}{\sigma^2_b}\ln(1+\frac{\sigma^2_b N_s}{N_b(N_b+\sigma^2_b)})         \right) \right]^{1/2}\,,
\end{equation}
with $N_s$ and $N_b$ being the number of signal and background events, respectively, and where $\sigma_b$ characterizes the uncertainty in the background events chosen to be the conservative value of $10\%$ \cite{Arganda:2022idg}. In this approximation, one expects to exclude regions with a total significance of $Z_A > 2$.

%%%=======================================%%%%
\subsection{The influence of cross-attention}
%%%=======================================%%%%

\begin{figure}[h!]
    \includegraphics[scale=0.38]{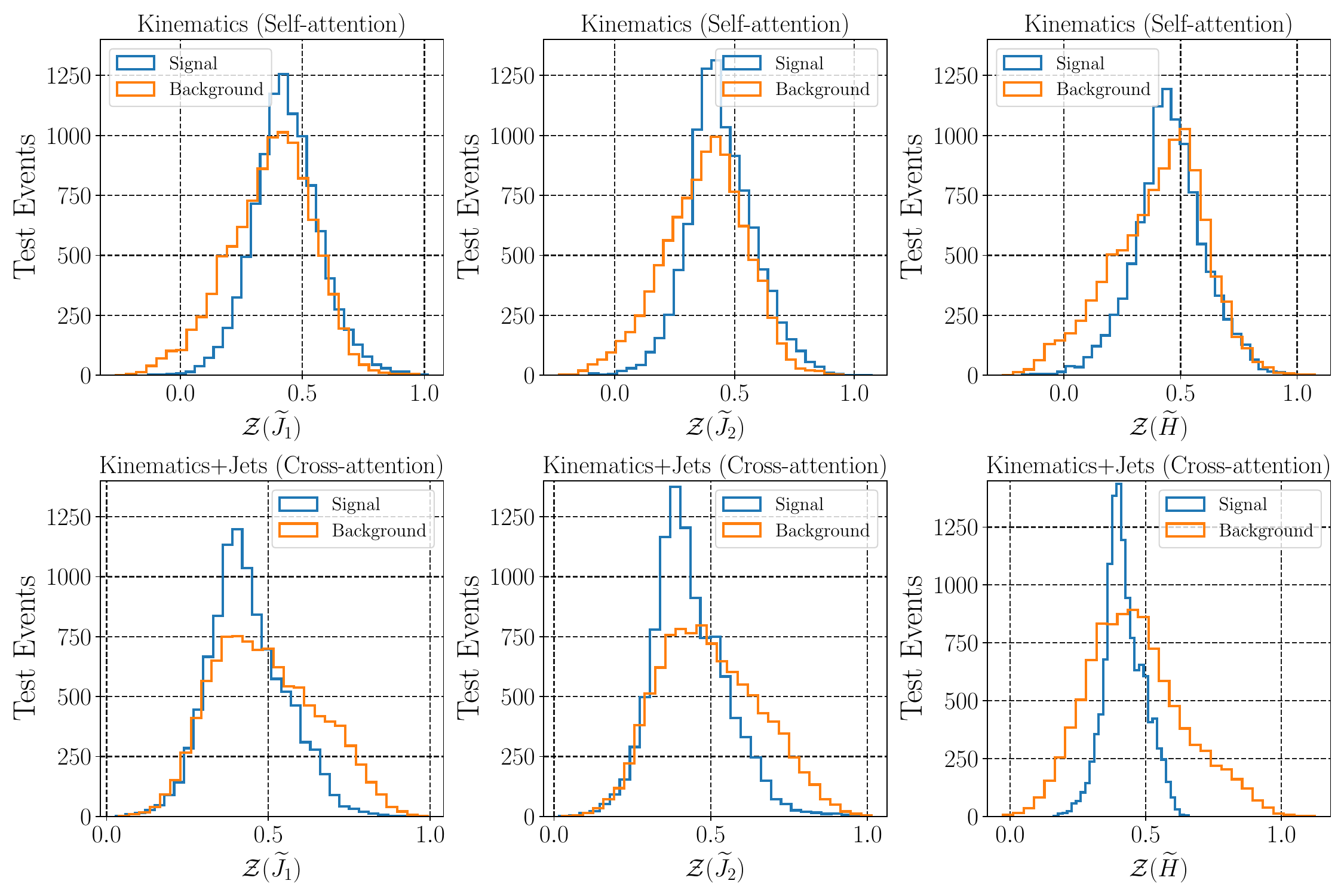}
    \caption{Top: output of the self-attention layer when trained on kinematics only. Bottom: output of the cross-attention layer when trained on kinematics and jet information. Attention output has the dimensions of (reconstructed particles $\times$ features), and for both plots we use 10000 test events and average over the features for the background and the signal point with $m_H=1$ TeV. $\widetilde{J_1}, \widetilde{J_2}$ and $\widetilde{H}$ represent the transformed particles as in equation \ref{eq:scaled_input}.}
    \label{fig:att_output}
\end{figure}

To evaluate the impact of the cross-attention layer on the classification performance, Fig.~\ref{fig:att_output} presents the attention output, as defined in Equation \ref{eq:attoutput}, for both the multi-modal transformer with cross-attention trained on kinematics plus jet constituents and the transformer network trained on kinematics only. In both networks, the attention output has dimensions of $(3, 6)$, where $3$ represents the reconstructed particles (leading, second-leading jet, and heavy Higgs), and the last dimension accounts for the utilized features. We stress that in the network structure shown in Fig. \ref{fig:feyn}, we adjust the Query matrix to be the output of the transformer layers for the jet constituents. In contrast, the Key and Value matrices are the output from the transformer layers for the kinematics. Accordingly, the output of the cross-attention layers has the same dimensions as the kinematics dataset. In principle, we have the freedom to choose whether to add the jet information to the kinematics by fixing the assigned Query, Key, and Value matrices, but we opted to incorporate the jet information into the high-level kinematics.

Fig.~\ref{fig:att_output} displays the distributions of the attention output for each transformed particle individually and averaged over the used features. 
 The top row shows the attention output for signal and background events using a transformer encoder trained on kinematics only. 

Conversely, when the information of the jet constituents is included using the cross-attention layer, the attention output distributions for background events are broader, and the signal distributions are narrower.  The fact that background jets lack a multi-prong structure with broader soft radiations influences the attention output for background events, increasing the output variations in the feature space.

\begin{figure}[h!]
    \includegraphics[scale=0.193]{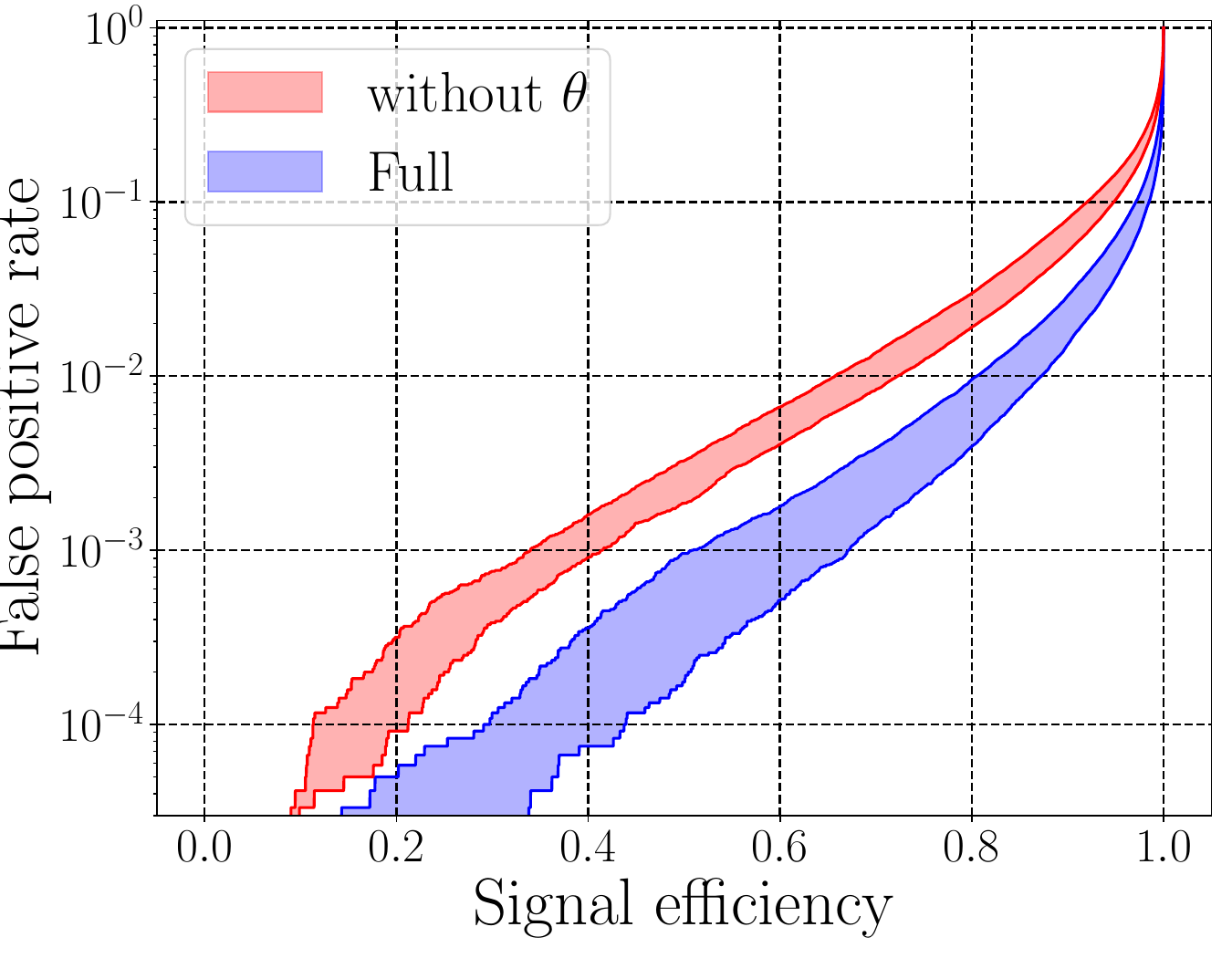}\includegraphics[scale=0.23]{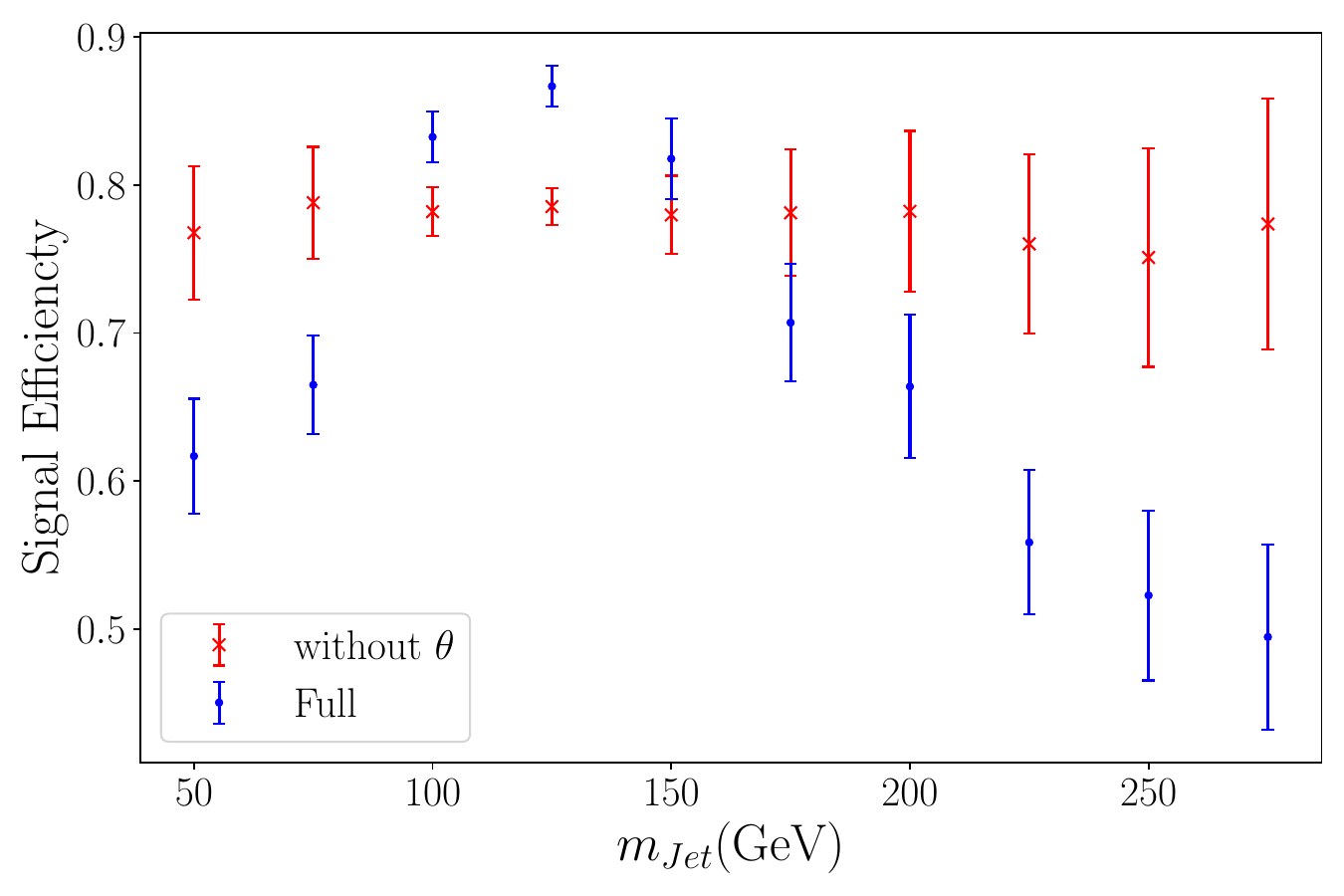}\includegraphics[scale=0.23]{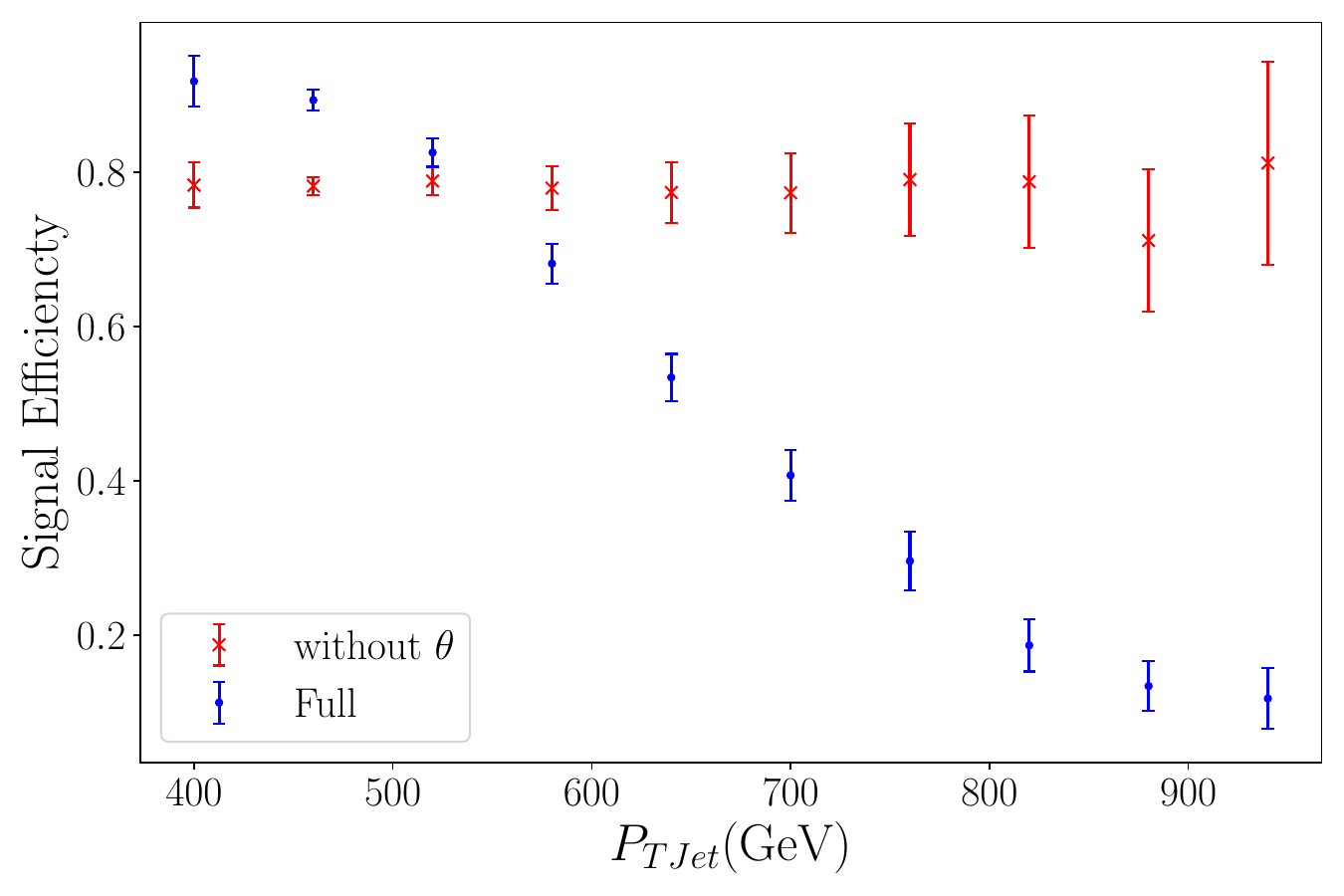}
    \caption{Left: The ROC curve and error band of the full model using $\theta$ input (blue) and the model without $\theta$ input (red).  The band indicates the max and min of the 5 independent training. The ROC is obtained by using 20,000 signal and background testing events. The error is estimated as in Fig. 6.  Middle(Right): plot shows the signal efficiency as varying $m_{J_1}$($p_{TJ_1}$) for the best training results. The ratio is calculated with the score cut of 80\% of the signal efficiency for 20,000 signal samples. The efficiency  (without) using $\theta$ is shown by blue(red) bars indicating statistical errors without taking into training errors. The acceptance of the full model is higher than without $\theta$ input at $m_{J_1}\sim m_h$ and $p_{J_1}\sim m_H/2$. }
    \label{fig:theta}
\end{figure}

{Finally, we include, alongside the described kinematical information, also the rotation angle $\theta$  aligning the fat jet axis to the $\phi$ direction after shifting the jet $\eta$ and $\phi$ to the center of the $\eta-\phi$ plane. 
This information allows the network to reconstruct the full events and access the correlation of the jet shape to the other fat jet and the beam axis. 

In Table \ref{tab:tab_2},  we compare the AUC value of the network using Kinematical inputs(Kins), Kins + $\theta$, Kins+ jet substructure inputs (Jet str.),  Kins + Jet str.+ $\theta$. 
Adding $\theta$ to Kins improves AUC by 0.59 while adding $\theta$ to Kins+ Jet str. improves AUC by 1.45. 
This indicates the correlation between all inputs (Jet str. $\theta$, and Kins) is contributing to the signal and background classification. 

\begin{table}[h!]
\centering
\caption{Area Under the ROC (AUC) for the networks using Kinematics or Kinematics + Jet structure information with/without $\theta$.  }\vspace*{0.25cm}
\begin{tabular}{|c|c|c|c|}
\hline
Kinematics  & Kinematics + $ \theta$ & Jet str. + Kinematics & Jet str. + Kinematics + $\theta$ \\ \hline
$91.01\%$ & $91.60\%$ & $97.23\%$ &$98.68\%$ \\  \hline
\end{tabular}
\label{tab:tab_2}
\end{table}

In Fig. \ref{fig:theta} Left, we show the ROC curve of the network trained without the $\theta$ inputs (red) compared to the ROC curve of our coss-attention model (blue). The improvement in the background rejection is a factor of four for a signal efficiency of 80\%.  Therefore,  including $\theta$ results in a drastically increased performance.  In Fig.  \ref{fig:theta} Middle and Right, we show the efficiency in rejecting background for the model with/without $\theta$ inputs. 
The model with $\theta$ has higher efficiency at $m_{J_1}\sim m_h$ and $p_T\sim \frac{m_H}{2}$. In short, the model can focus more on the $H\rightarrow h h $ kinematics with $\theta$ inputs. We also looked for simple correlations among $\theta$ and the other kinematical variables, such as $\eta_J$ $\phi_J$, but did not find any apparent ones contributing to the selection improvement, consistent with insignificant improvement by adding $\theta$ to the model using Kins. The correlations within  the internal structures of the jet will be investigated in future publications. 
 }

%\textcolor{blue}{In table \ref{tab:tab_2} we show the AUC for ....}

%%%-----------------------------------
\section{Interpretation of the transformer encoder results}
%%%------------------------------------
In the following section, we discuss additional methods to interpret and analyze the results of the transformer encoder with cross-attention, which performs best in Fig. 6 
The interpretation methods are generic and can be further applied to other networks to interpret their results.
As attention-based transformer models excel in capturing intricate spatial relationships and global context within data, their interpretability becomes paramount.  
Interpretation methods for attention-based transformers aim to elucidate the visual cues, features, and regions that contribute significantly to the model's predictions. 
Common Interpretation Methods  are 
\begin{itemize}
    \item {\bf Attention Maps:} Attention maps visualize the focus of the model by highlighting the particles in the cloud that receive higher attention. These maps provide a direct view into which particles are considered most relevant for prediction, facilitating an intuitive understanding of the model's decision-making process. 
    
     \item {\bf Grad-CAM}  It generates class-specific activation maps by weighting the gradients of the predicted class score with respect to the final transformer layer \cite{selvaraju2017grad}. This technique highlights the regions in the feature space that are crucial for the model's classification decision and thus can provide a geometrical interpretation, $\eta-\phi$ plane,  of the learned information by the network. 
     
     \item{\bf Saliency Maps} Saliency maps for transformer models are a form of interpretability technique used to understand and visualize the importance of different parts of the input sequence concerning the model's predictions. Saliency maps highlight the regions of the input that most significantly influence the model's output, providing insights into the model's decision-making process \cite{huang2022ssit,duong2022put,lu2023saliency}. By examining the saliency map, users can gain insights into which parts of the input sequence are crucial for the model's predictions.
     
      \item{\bf Layer-wise Relevance Propagation (LRP)} The primary goal of LRP is to assign relevance scores to input features, indicating their contribution to the model's output \cite{binder2016layer}. However, it's worth noting that LRP has limitations, and its effectiveness can vary depending on the specific neural network architecture and the nature of the task. Different variants of LRP have been proposed to address specific challenges and improve its applicability to various models. 
\end{itemize}

The interpretation of attention-based transformer models is pivotal for unlocking their full potential and ensuring their responsible deployment in real-world applications. Among all the mentioned methods, we adopt the attention maps and Grad-CAM to interpret the learned information using the transformer model.
%%--------------------------------------
\subsection{Attention maps}
%%%---------------
Attention maps serve as a bridge between the abstract nature of neural network computations and the desired interpretability. These maps visualize the attention scores assigned to each particle token in the input sequence, providing a representation of where the model focuses its attention during processing. 
\begin{figure}[th!]
    \includegraphics[scale=0.28]{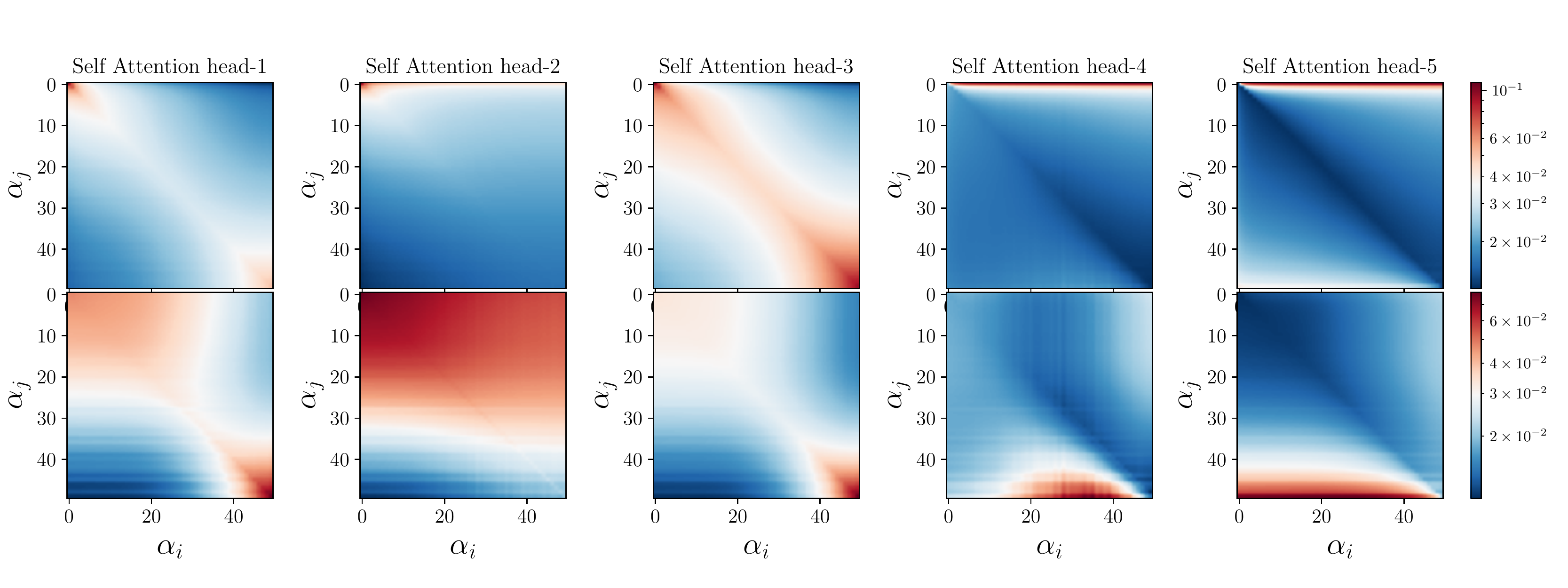}
    \caption{Attention maps of the last self-attention transformer layer, which processes the jet substructure for the signal (top) and backgrounds (bottom) for a 120K test event.}
    \label{fig:attention_jet}
\end{figure}

\begin{figure}[th!]
    \includegraphics[scale=0.32]{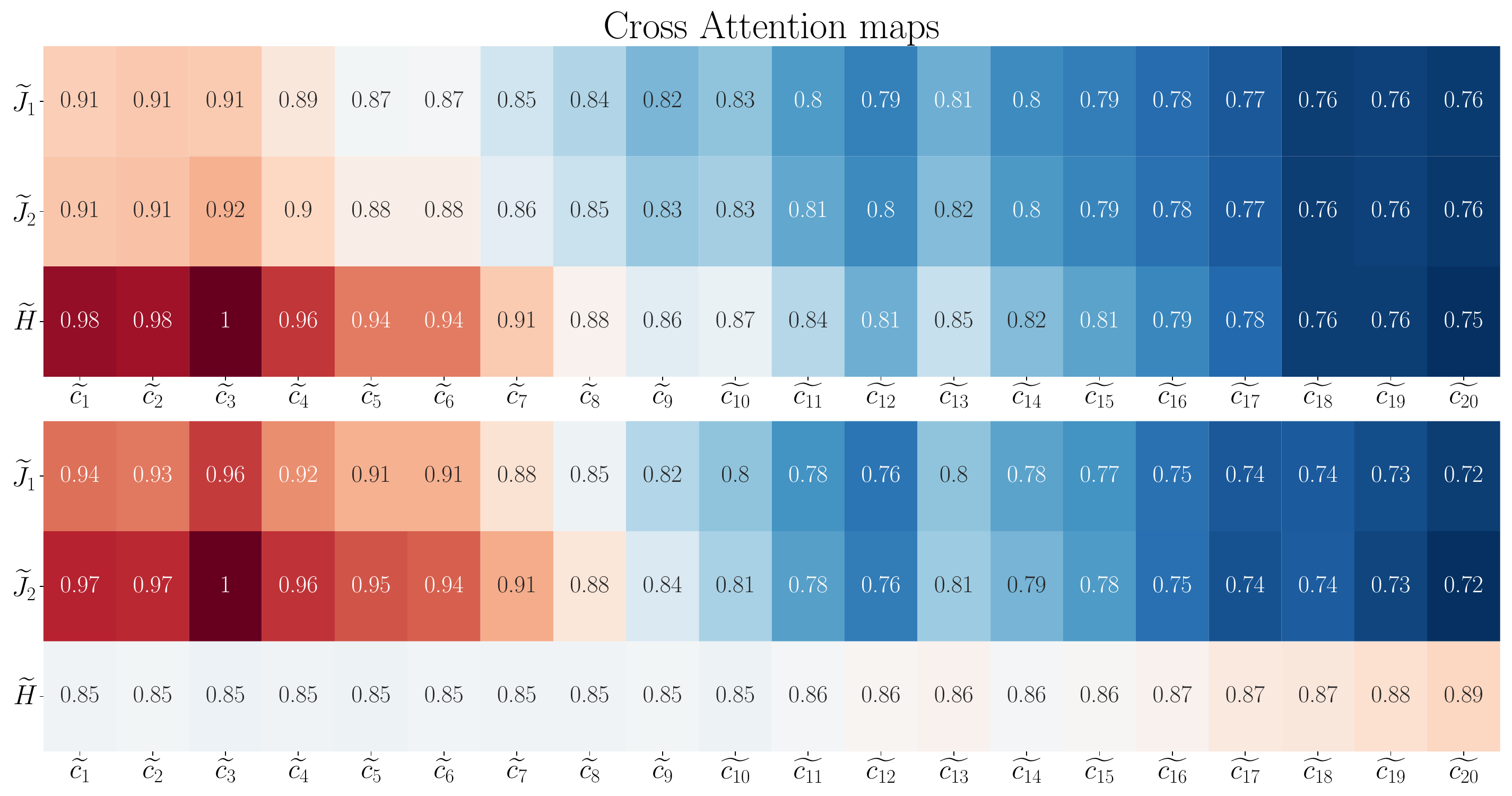}
    \caption{Cross-attention maps of the cross-attention transformer layer averaged over the  8 cross-attention heads, which processes the jet substructure and the event kinematics for the signal (top) and backgrounds (bottom) for a 120K test event. The X-axis shows the attention score for the first transformed $20^{\text{th}}$ jet contents while the Y-axis shows the attention score for the  transformed reconstructed final state particles.}
    \label{fig:attention_kin}
\end{figure}
The analysis of the attention maps highlights the particle tokens that receive higher attention scores, indicating their significance in the model's decision. Also, it reveals how particle tokens relate to each other. For instance, it highlights the information extracted from the jet constituents relevant to the reconstructed objects. Importantly, examining attention maps can pinpoint areas where the model might struggle or make mistakes.

In this context, we utilize attention maps to analyze the acquired information from the last transformer layer of the jet constituents. Our focus centers on the output of the network shown in Fig. \ref{fig:network}. We begin by examining the attention maps of the Add() layer, which contains information about the jet substructure. In this case, the attention maps denoted as $\alpha_{ij}$ in Equation \ref{eq:alpha}, have dimensions of $(n_\text{heads}, 50, 50)$, where $50$ represents the number of constituents in the jet, and $n_\text{heads}$ denotes the number of self-attention heads.  We take $n_\text{heads}=5$ , see Appendix \ref{appendix:1}) for detail.

Fig. \ref{fig:attention_jet} displays the values of the attention maps for each attention head individually, with signal events in the top row and background events in the bottom row. Given that jet constituents are originally ordered by their momentum, the X and Y axes ticks represent the attention values of the jet constituents in descending order (where the zero tick represents the leading transformed jet constituent particle). The attention map values reveal that the model concentrates on the leading and second-leading jet constituents to identify events as signal-like, particularly evident in attention heads $1, 2, 4, $ and $5$.  While the transformer layers intermingle particle and feature tokens, the skip connection still preserves the order of the attention output in relation to the original input data. In fact, this reflects the efficiency of the network to capture the two-prong structure of the signal events. On the other hand, the network assigns high attention to the wide momentum orders of the jet constituents when the network identifies the input as a background event. Conversely, the network assigns high attention to a broad momentum range of jet constituents when identifying the input as a background event. The attention maps for background events exhibit significant agreement with the jet substructure of the background events presented in Fig. \ref{fig:jet_images}. To this end, through an analysis of the attention scores from the last transformer layer of the jet constituents, we confirm that the transformer model adeptly extracts the correct multi-prong structure of signal events. Meanwhile, for background events dominated by QCD processes, the model exhibits high attention across a wide momentum range of jet constituents.

The attention maps of the cross-attention layer illustrate the attention scores between the jet constituents and the reconstructed particles, including the leading and second-leading jets and the heavy Higgs. The dimension of the attention score in the cross-attention layer is $(n_{heads}, 3, 50)$, where $3$ represents the number of reconstructed particles, $50$ is the number of jet constituents, and $n_{heads}$ is the number of cross-attention heads, set at $8$. Fig. \ref{fig:attention_kin} displays the cross-attention maps for signal events (top) and background events (bottom), averaged over the used cross-attention heads.

The cross-attention maps for signal events exhibit a stronger correlation between the highest momentum jet constituents and the heavy Higgs. In contrast, the Heavy Higgs displays a flat attention pattern with jet constituents of different momenta. Indeed, the results from the cross-attention maps, along with the cross-attention output shown in Fig. \ref{fig:att_output}, provide a comprehensive overview of the impact of the cross-attention layer. This layer effectively assigns information from the jet constituents to the kinematics of the reconstructed particles to enhance the classification performance.

%%--------------------------------------
\subsection{Grad-CAM}
%%%---------------

Grad-CAM is a technique designed to visualize and interpret the decisions made by DNN models. It builds upon the idea of class activation maps (CAMs)\cite{cherepanov2022visualization,zhou2016learning} but extends it to models with arbitrary architectures. The primary objective of Grad-CAM is to highlight the important regions in a transformed input features space, $\widetilde{\eta}-\widetilde{\phi}$ plane,  that contribute to the prediction of a specific class \cite{chefer2021transformer}.

Let $F_k(\widetilde{\eta},\widetilde{\phi})$ represent the activation of the final transformer layer for the $k^{\text{th}}$ event. The gradient of the predicted class score ($Y_c$) with respect to the activation output is computed as:
\begin{equation}
\frac{\partial Y_c}{\partial F_k}
\end{equation}
This gradient is then globally averaged to obtain the importance weights ($\alpha$) as
\begin{equation}
\alpha_k(\widetilde{\eta},\widetilde{\phi}) = \frac{1}{Z} \sum \frac{\partial Y_c}{\partial F_k(\widetilde{\eta},\widetilde{\phi},\widetilde{p_T})}
\end{equation}
where $Z$ is the size of the feature activations, and the sum runs over the jet constituents. $\widetilde{\eta}$, $\widetilde{\phi}$ and $\widetilde{p_T}$ are the transformed features. 
%The final Grad-CAM heatmap  is computed as a weighted  of the summed  gradients as
The final Grad-CAM heatmap is a weighted sum of gradients as 
\begin{equation}
\text{Grad-CAM}(\widetilde{\eta},\widetilde{\phi}) = \frac{1}{k} \sum_k \alpha_k(\widetilde{\eta},\widetilde{\phi}) F_k(\widetilde{\eta}, \widetilde{\phi},\widetilde{p_T})
\end{equation}

This heatmap highlights the regions of the input image that contribute the most to the prediction of the target class.

\begin{figure}[th!]
    \includegraphics[width=15cm,height=10cm]{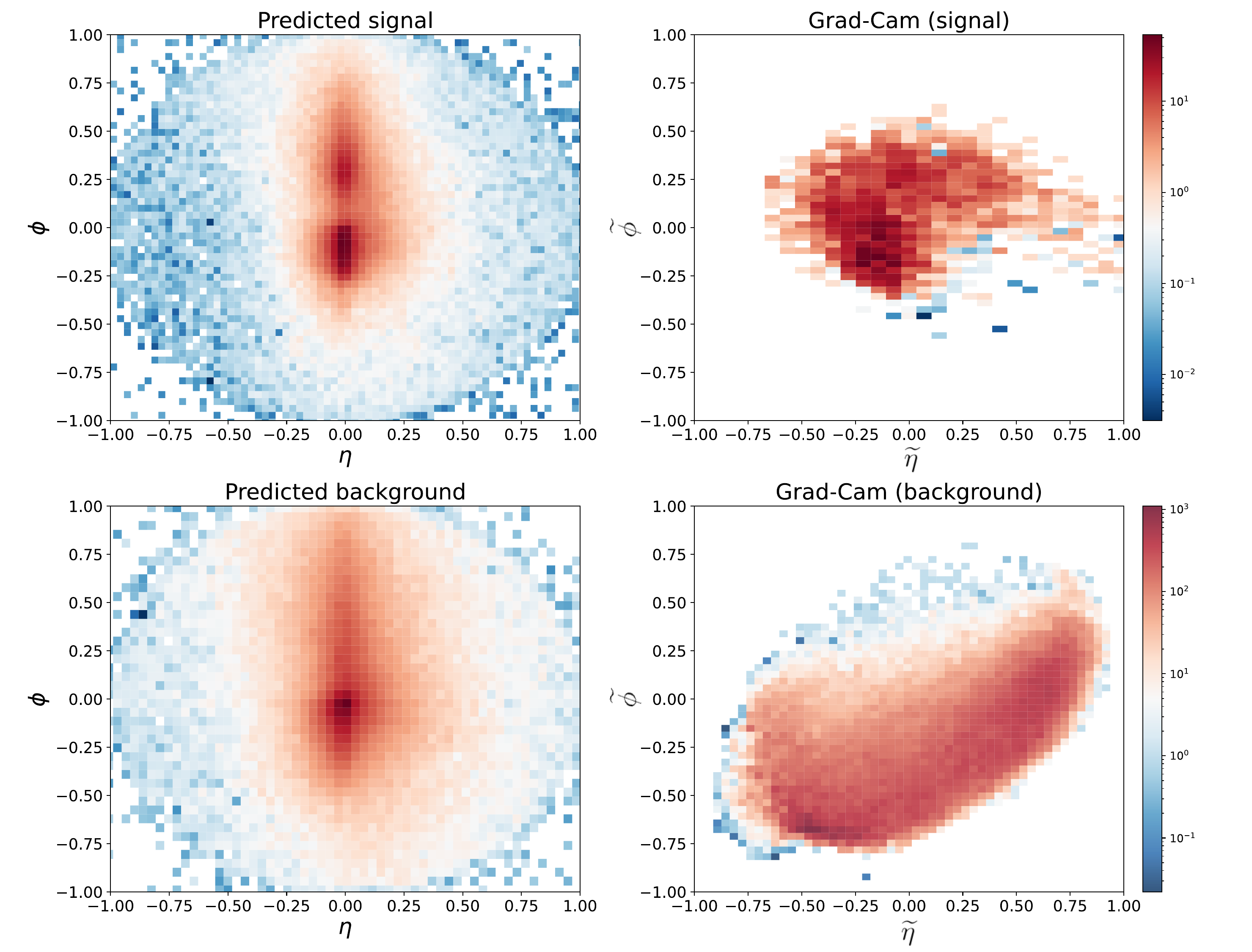}
    \caption{Grad-CAM results for 5000 test events of the transformer model with cross-attention. Left: $p_T$ distribution of the jet constituents when events are predicted as signal events (top) and background events (bottom) in $\widetilde{\eta}$-$\widetilde{\phi}$ plane. Note that the asymmetric pattern is due to the flipping transformation in the pre-processing steps in which all constituents with larger momentum are reflected in the positive $\eta$ direction. Right: heat-map of the Grad-CAM results in $\widetilde{\eta}$-$\widetilde{\phi}$ plane. }
    \label{fig:gradcam}
\end{figure}

In general, it operates by utilizing the gradient information flowing into the final transformer layer in the following way: During the forward pass, the neural network processes the input particle cloud, and the activations of the final transfomer layer are obtained. The gradients of the predicted class score with respect to the final transformer layer activation are computed during the backward pass. The gradients are then used to calculate the importance of the activation map. These importance scores are essentially the weights assigned to each spatial location of the final transformer layer. The weighted sum of the particle tokens is computed, creating the Grad-CAM. This map highlights the regions that contributed the most to the final prediction. Additionally, upsampling is often employed to match the Grad-CAM dimensions with the original input features. 

To visualize the geometrical interpretation of the learned information from the jet constituents, we utilize Grad-CAM on the final self-attention layer of the jets, specifically, the Add() Layer depicted in Fig. \ref{fig:network}. The results are shown in Fig. \ref{fig:gradcam} for 5000 test images. The left panel illustrates the $p_T$ distribution of the predicted events as signal (top) or background (bottom). Signal events are considered for the benchmark point with $m_H=1$ TeV. The right panel displays the heat map of the Grad-CAM output for the predicted signal (top) and the predicted background (bottom).

The visualization of the heat map clarifies that the transformer model focuses on the two-prong structure to classify the input event as a signal. On the other hand, it relies on the soft-radiation pattern to classify the input event as a background event. Interestingly, we found that the result highlights that the model focuses on the positive $\eta$ direction to make predictions, which is due to the flipping transformation done in the pre-processing step.

While Grad-CAM has the power to explain the considered regions in the feature space for the network predictions, one of its drawbacks is that it relies on gradient information from the final transformer layer. In cases where global context is crucial for decision-making, Grad-CAM may not capture long-range dependencies effectively. Moreover, Grad-CAM might be sensitive to small changes in the input, potentially making it less robust in the presence of adversarial examples.

%%--------------------------------------
\section{Conclusion}
%%%---------------

In conclusion, this paper introduces an innovative method for enhancing event classification by effectively incorporating information from both global kinematics and substructure of jets in an event. 
Conventional approaches,  using simple concatenation to combine the event information, have limitations, especially for scenarios where kinematical structures dominate. 

Specifically, the proposed method utilizes a transformer encoder with cross-attention layers, enabling the extraction of different scale information from both global kinematics and jet substructure. 
The results demonstrate a substantial improvement in classification performance compared to traditional concatenation methods.
Indeed, the analysis of the learned information, conducted through attention maps and a Grad-CAM algorithm for visual representation, provides valuable insights into the model focus on important particles and geometric regions in the transformed $\widetilde{\eta}-\widetilde{\phi}$  plane that is crucial for event classification.

We have validated 
this approach by focusing on the dominant decay channel, i.e., into four $b$-jets, of SM-like Higgs boson pairs produced
in the resonant decay of a heavier CP-even Higgs state,  
at the HL-LHC. This challenging scenario involves merging would-be slim $b$-jets into a fatjet, due to the boosted nature of the lighter Higgs states so that the possibility of accessing partonic dynamics is apparently lost at the detector level. Furthermore, this occurs in an environment rich in tracks and calorimetric information not directly pertaining to the hard scattering sought, as typical of this CERN machine upgrade. Therefore, all these aspects add complexity to the classification task. Despite these challenges, the proposed method effectively addresses the intricacies of the final state in the detectors, ultimately outperforming mainstream signal selection
procedures, whether based solely on kinematical analysis or less advanced ML tools.
In the broader context, this research contributes to utilizing advanced jet identification techniques for global event reconstruction towards the understanding of collision events consisting of dynamics acting at various physics scales. 
Thus, the proposed method offers a promising avenue for improving the accuracy and efficiency of event classification in potentially many more complex scenarios encountered in high energy physics experiments.
%%%%%%%%%%%%%%%%%%%%%%%%%%%%

\subsection*{Acknowledgments}
The work of SM is supported in part through the NExT Institute, the Knut and Alice Wallenberg Foundation under the Grant No. KAW 2017.0100 (SHIFT) and the STFC Consolidated Grant No. ST/L000296/1. AH and MN are funded by grant number 22H05113, ``Foundation of Machine Learning Physics'', Grant in Aid for Transformative Research Areas and 22K03626, Grant-in-Aid for Scientific Research (C).
\appendix

%%%%%%%%%%%%%%%%%%%%%%%%%%%%%%
\section{Networks structure}
\label{appendix:1}
%%%%%%%%%%%%%%%%%%%%%%%%%%%%%
In this study, we employed four transformer encoders with distinct configurations to analyze various datasets. For all networks, we configured an output layer with two neurons and applied softmax activation. Additionally, we utilized the Adam optimizer \cite{kingma2014adam} to minimize the sparse categorical cross-entropy loss function \cite{terven2023loss}, setting the learning rate at $0.005$. Our training dataset comprised one million samples, with 20,000 allocated for validation and 100,000 for testing. The training batches were adjusted to a size of 500.

Following data preprocessing, we obtained three datasets: one for the leading jet contents with dimensions of $(50,4)$, where $50$ represents the number of jet constituents and $4$ denotes the considered features... We also use the same information for the second leading jet contents. 
%mirroring the dimensions of the leading jet dataset; 
The last dataset for event kinematics with dimensions of $(3,6)$, where $3$ corresponds to the leading jet, the second leading jet, and the (reconstructed) heavy Higgs boson. The structure of the different networks is the following:
\begin{itemize}
    \item A two-stream self-attention transformer encoder is employed for jet substructure. 
    The network takes two separate data sets for the leading and second-leading jet constituents as input, processed through two distinct transformer layers. 
    Each transformer layer is repeated three times. These transformer layers consist of five self-attention heads operating in parallel. 
    The output from the attention heads is then integrated with the original input data via a skip connection layer \cite{lai2022rethinking}. 
    The resulting output from the skip connection is flattened and forwarded to two fully connected layers with 128 and 4 neurons, respectively, using the GELU activation function \cite{hendrycks2016gaussian}. 
    The output from the final fully connected layer is subsequently combined with the self-attention output through a second skip connection layer\footnote{Skip connection is of the utmost importance to stabilize the gradient flow of the model.}. 
    The final output of the transformer layer undergoes %normalization using 
    a normalization layer and has the same dimension as the input dataset. The normalized output from each transformer layer is combined through an addition layer. This output then passes through a Multi-Layer Perceptron (MLP) comprising two fully connected layers with dimensions 128 and 64, employing the GELU activation function. Following each fully connected layer, a dropout layer with a dropout rate of $20\%$ is applied. The output is then passed to the output layer for classification. The model is trained for 30 epochs with batch of size $500$ in $1421$ seconds.

    \item A single-stream self-attention transformer encoder is employed for kinematics analysis. The network exclusively utilizes the kinematics dataset as input. To achieve this, we adopt the identical structure of the self-attention transformer encoder designed for jet substructure, but with a singular stream. The model is trained for 30 epochs with batch of size $500$ in $1390$ seconds.

    \item A three-stream transformer encoder is employed to analyze the leading and subleading jet constituents and the reconstructed kinematics. In this approach, we adjust the transformer layers for the leading and subleading jets from the first network, while the transformer layers for the kinematics are adapted from the latter network. The output of the self-attention transformer encoder layers for jet constituents is added via an addition layer. The resulting output from the addition layer, along with the output from the self-attention transformer layers of the kinematics, is then fed to a cross-attention transformer layer. This cross-attention transformer layer is repeated twice, and the output has the same dimensions as the input kinematics dataset, i.e., $(3,6)$. Subsequently, this output passes through a MLP consisting of two fully connected layers with dimensions 128 and 64, utilizing the GELU activation function. After each fully connected layer, a dropout layer with a dropout rate of $20\%$ is applied. The resulting output is then forwarded to the output layer for classification. The model is trained for 30 epochs with batch of size $500$ in $1576$ seconds.

    \item The final network is configured to mirror the three-stream transformer encoder, with the only modification being the substitution of the cross-attention transformer layers with a single concatenation layer. The model is trained for 30 epochs with batch of size $500$ in $1282$ seconds.
\end{itemize}
For training of all models, we use two NVIDIA RTX 6000 GPU cards using Tensorflow mirror strategy with the utilization of $80\%$ and $30\%$ for the first and second cards, respectively, and memory consumption of $96\%$ (48 GB) of both cards.

%%%%%%%%%%%%%%%%%%%%
\bibliographystyle{unsrt}
\bibliography{biblo}

\begin{thebibliography}{100}

\bibitem{Chakraborty:2019imr}
Amit Chakraborty, Sung~Hak Lim, and Mihoko~M. Nojiri.
\newblock {Interpretable deep learning for two-prong jet classification with
  jet spectra}.
\newblock {\em JHEP}, 07:135, 2019.

\bibitem{Chung:2020ysf}
Yi-Lun Chung, Shih-Chieh Hsu, and Benjamin Nachman.
\newblock {Disentangling Boosted Higgs Boson Production Modes with Machine
  Learning}.
\newblock {\em JINST}, 16:P07002, 2021.

\bibitem{Guo:2020vvt}
Jun Guo, Jinmian Li, Tianjun Li, and Rao Zhang.
\newblock {Boosted Higgs boson jet reconstruction via a graph neural network}.
\newblock {\em Phys. Rev. D}, 103(11):116025, 2021.

\bibitem{Khosa:2021cyk}
Charanjit~K. Khosa and Simone Marzani.
\newblock {Higgs boson tagging with the Lund jet plane}.
\newblock {\em Phys. Rev. D}, 104(5):055043, 2021.

\bibitem{Datta:2019ndh}
Kaustuv Datta, Andrew Larkoski, and Benjamin Nachman.
\newblock {Automating the Construction of Jet Observables with Machine
  Learning}.
\newblock {\em Phys. Rev. D}, 100(9):095016, 2019.

\bibitem{Cogollo:2020afo}
D.~Cogollo, F.~F. Freitas, C.~A. de~S.~Pires, Yohan~M. Oviedo-Torres, and
  P.~Vasconcelos.
\newblock {Deep learning analysis of the inverse seesaw in a 3-3-1 model at the
  LHC}.
\newblock {\em Phys. Lett. B}, 811:135931, 2020.

\bibitem{Grossi:2020orx}
M.~Grossi, J.~Novak, B.~Kersevan, and D.~Rebuzzi.
\newblock {Comparing traditional and deep-learning techniques of kinematic
  reconstruction for polarization discrimination in vector boson scattering}.
\newblock {\em Eur. Phys. J. C}, 80(12):1144, 2020.

\bibitem{Ngairangbam:2020ksz}
Vishal~S. Ngairangbam, Akanksha Bhardwaj, Partha Konar, and Aruna~Kumar Nayak.
\newblock {Invisible Higgs search through Vector Boson Fusion: A deep learning
  approach}.
\newblock {\em Eur. Phys. J. C}, 80(11):1055, 2020.

\bibitem{Englert:2020ntw}
Christoph Englert, Malcolm Fairbairn, Michael Spannowsky, Panagiotis Stylianou,
  and Sreedevi Varma.
\newblock {Sensing Higgs boson cascade decays through memory}.
\newblock {\em Phys. Rev. D}, 102(9):095027, 2020.

\bibitem{Freitas:2020ttd}
Felipe~F. Freitas, Jo\~ao Gon\c{c}alves, Ant\'onio~P. Morais, and Roman
  Pasechnik.
\newblock {Phenomenology of vector-like leptons with Deep Learning at the Large
  Hadron Collider}.
\newblock {\em JHEP}, 01:076, 2021.

\bibitem{Stakia:2021pvp}
Anna Stakia et~al.
\newblock {Advances in Multi-Variate Analysis Methods for New Physics Searches
  at the Large Hadron Collider}.
\newblock {\em Rev. Phys.}, 7:100063, 2021.

\bibitem{Jorge:2021vpo}
Fraga Jorge, Rodriguez Ronald, Solano Jesus, Molano Juan, and Avila Carlos.
\newblock {Top squark signal significance enhancement by different machine
  learning algorithms}.
\newblock {\em Int. J. Mod. Phys. A}, 37(31n32):2250197, 2022.

\bibitem{Ren:2021prq}
Jie Ren, Daohan Wang, Lei Wu, Jin~Min Yang, and Mengchao Zhang.
\newblock {Detecting an axion-like particle with machine learning at the LHC}.
\newblock {\em JHEP}, 11:138, 2021.

\bibitem{Alvestad:2021sje}
Daniel Alvestad, Nikolai Fomin, J\"orn Kersten, Steffen Maeland, and Inga
  Str\"umke.
\newblock {Beyond cuts in small signal scenarios: Enhanced sneutrino
  detectability using machine learning}.
\newblock {\em Eur. Phys. J. C}, 83(5):379, 2023.

\bibitem{Jung:2021tym}
Sunghoon Jung, Zhen Liu, Lian-Tao Wang, and Ke-Pan Xie.
\newblock {Probing Higgs boson exotic decays at the LHC with machine learning}.
\newblock {\em Phys. Rev. D}, 105(3):035008, 2022.

\bibitem{Drees:2021oew}
Manuel Drees, Meng Shi, and Zhongyi Zhang.
\newblock {Machine Learning Optimized Search for the $Z'$ from
  $U(1)_{L_\mu-L_\tau}$ at the LHC}.
\newblock 9 2021.

\bibitem{Cornell:2021gut}
Alan~S. Cornell, Wesley Doorsamy, Benjamin Fuks, Gerhard Harmsen, and Lara
  Mason.
\newblock {Boosted decision trees in the era of new physics: a smuon analysis
  case study}.
\newblock {\em JHEP}, 04:015, 2022.

\bibitem{Vidal:2021oed}
Xabier~Cid Vidal, Lorena~Dieste Maro\~nas, and \'Alvaro~D\'osil Su\'arez.
\newblock {How to Use Machine Learning to Improve the Discrimination between
  Signal and Background at Particle Colliders}.
\newblock {\em Appl. Sciences}, 11(22):11076, 2021.

\bibitem{Lin:2018cin}
Joshua Lin, Marat Freytsis, Ian Moult, and Benjamin Nachman.
\newblock {Boosting $H\to b\bar b$ with Machine Learning}.
\newblock {\em JHEP}, 10:101, 2018.

\bibitem{Moreno:2019neq}
Eric~A. Moreno, Thong~Q. Nguyen, Jean-Roch Vlimant, Olmo Cerri, Harvey~B.
  Newman, Avikar Periwal, Maria Spiropulu, Javier~M. Duarte, and Maurizio
  Pierini.
\newblock {Interaction networks for the identification of boosted $H
  \rightarrow b\overline{b}$ decays}.
\newblock {\em Phys. Rev. D}, 102(1):012010, 2020.

\bibitem{Chung:2022kjp}
Yi-Lun Chung, Kingman Cheung, and Shih-Chieh Hsu.
\newblock {Sensitivity of two-Higgs-doublet models on Higgs-pair production via
  bb\textasciimacron{}bb\textasciimacron{} final state}.
\newblock {\em Phys. Rev. D}, 106(9):095015, 2022.

\bibitem{Kim:2019wns}
Jeong~Han Kim, Minho Kim, Kyoungchul Kong, Konstantin~T. Matchev, and Myeonghun
  Park.
\newblock {Portraying Double Higgs at the Large Hadron Collider}.
\newblock {\em JHEP}, 09:047, 2019.

\bibitem{Huang:2022rne}
Li~Huang, Su-beom Kang, Jeong~Han Kim, Kyoungchul Kong, and Jun~Seung Pi.
\newblock {Portraying double Higgs at the Large Hadron Collider II}.
\newblock {\em JHEP}, 08:114, 2022.

\bibitem{Esmail:2023axd}
W.~Esmail, A.~Hammad, and S.~Moretti.
\newblock {Sharpening the A \textrightarrow{} Z$^{(*)}$h signature of the
  Type-II 2HDM at the LHC through advanced Machine Learning}.
\newblock {\em JHEP}, 11:020, 2023.

\bibitem{Ban:2023jfo}
Kayoung Ban, Kyoungchul Kong, Myeonghun Park, and Seong~Chan Park.
\newblock {Exploring the Synergy of Kinematics and Dynamics for Collider
  Physics}.
\newblock 11 2023.

\bibitem{Chakraborty:2020vwj}
Amit Chakraborty, Srinandan Dasmahapatra, Henry Day-Hall, Billy Ford, Shubhani
  Jain, Stefano Moretti, Emmanuel Olaiya, and Claire Shepherd-Themistocleous.
\newblock {Revisiting jet clustering algorithms for new Higgs Boson searches in
  hadronic final states}.
\newblock {\em Eur. Phys. J. C}, 82(4):346, 2022.

\bibitem{Chakraborty:2022lcj}
Amit Chakraborty, Srinandan Dasmahapatra, Henry Day-Hall, Billy Ford, Shubhani
  Jain, Stefano Moretti, Emmanuel Olaiya, and Claire Shepherd-Themistocleous.
\newblock {Re-evaluating Jet Reconstruction Techniques for New Higgs Boson
  Searches}.
\newblock {\em PoS}, ICHEP2022:503, 2022.

\bibitem{Chakraborty:2023hrk}
Amit Chakraborty, Srinandan Dasmahapatra, Henry Day-Hall, Billy Ford, Shubhani
  Jain, and Stefano Moretti.
\newblock {Fat b-jet analyses using old and new clustering algorithms in new
  Higgs boson searches at the LHC}.
\newblock {\em Eur. Phys. J. C}, 83(4):347, 2023.

\bibitem{Cerro:2022rpf}
Giorgio Cerro, S.~Dasmahapatra, H.~A. Day-Hall, B.~Ford, S.~Jain, S.~Moretti,
  and C.~Shepherd-Themistocleous.
\newblock {Spectral clustering for jet reconstruction}.
\newblock {\em PoS}, ICHEP2022:771, 11 2022.

\bibitem{vaswani2017attention}
Ashish Vaswani, Noam Shazeer, Niki Parmar, Jakob Uszkoreit, Llion Jones,
  Aidan~N Gomez, {\L}ukasz Kaiser, and Illia Polosukhin.
\newblock Attention is all you need.
\newblock {\em Advances in neural information processing systems}, 30, 2017.

\bibitem{Kach:2022uzq}
Benno K\"ach, Dirk Kr\"ucker, and Isabell Melzer-Pellmann.
\newblock {Point Cloud Generation using Transformer Encoders and Normalising
  Flows}.
\newblock 11 2022.

\bibitem{Finke:2023veq}
Thorben Finke, Michael Kr\"amer, Alexander M\"uck, and Jan T\"onshoff.
\newblock {Learning the language of QCD jets with transformers}.
\newblock {\em JHEP}, 06:184, 2023.

\bibitem{Qu:2022mxj}
Huilin Qu, Congqiao Li, and Sitian Qian.
\newblock {Particle Transformer for Jet Tagging}.
\newblock 2 2022.

\bibitem{Komiske:2018cqr}
Patrick~T. Komiske, Eric~M. Metodiev, and Jesse Thaler.
\newblock {Energy Flow Networks: Deep Sets for Particle Jets}.
\newblock {\em JHEP}, 01:121, 2019.

\bibitem{Qu:2019gqs}
Huilin Qu and Loukas Gouskos.
\newblock {ParticleNet: Jet Tagging via Particle Clouds}.
\newblock {\em Phys. Rev. D}, 101(5):056019, 2020.

\bibitem{Branco:2011iw}
G.~C. Branco, P.~M. Ferreira, L.~Lavoura, M.~N. Rebelo, Marc Sher, and Joao~P.
  Silva.
\newblock {Theory and phenomenology of two-Higgs-doublet models}.
\newblock {\em Phys. Rept.}, 516:1--102, 2012.

\bibitem{Lee:1973iz}
T.~D. Lee.
\newblock {A Theory of Spontaneous T Violation}.
\newblock {\em Phys. Rev. D}, 8:1226--1239, 1973.

\bibitem{Glashow:1976nt}
Sheldon~L. Glashow and Steven Weinberg.
\newblock {Natural Conservation Laws for Neutral Currents}.
\newblock {\em Phys. Rev. D}, 15:1958, 1977.

\bibitem{Ginzburg:2004vp}
Ilya~F. Ginzburg and Maria Krawczyk.
\newblock {Symmetries of two Higgs doublet model and CP violation}.
\newblock {\em Phys. Rev. D}, 72:115013, 2005.

\bibitem{Antusch:2020ngh}
Stefan Antusch, Oliver Fischer, A.~Hammad, and Christiane Scherb.
\newblock {Testing CP Properties of Extra Higgs States at the HL-LHC}.
\newblock {\em JHEP}, 03:200, 2021.

\bibitem{Antusch:2021oit}
Stefan Antusch, Oliver Fischer, A.~Hammad, and Christiane Scherb.
\newblock {Explaining excesses in four-leptons at the LHC with a double peak
  from a CP violating Two Higgs Doublet Model}.
\newblock {\em JHEP}, 08:224, 2022.

\bibitem{Arhrib:2009hc}
Abdesslam Arhrib, Rachid Benbrik, Chuan-Hung Chen, Renato Guedes, and Rui
  Santos.
\newblock {Double Neutral Higgs production in the Two-Higgs doublet model at
  the LHC}.
\newblock {\em JHEP}, 08:035, 2009.

\bibitem{Hammad:2022wpq}
A.~Hammad, Myeonghun Park, Raymundo Ramos, and Pankaj Saha.
\newblock {Exploration of parameter spaces assisted by machine learning}.
\newblock {\em Comput. Phys. Commun.}, 293:108902, 2023.

\bibitem{ATLAS:2022hwc}
Georges Aad et~al.
\newblock {Search for resonant pair production of Higgs bosons in the
  $b\bar{b}b\bar{b}$ final state using $pp$ collisions at $\sqrt{s}$ = 13 TeV
  with the ATLAS detector}.
\newblock {\em Phys. Rev. D}, 105(9):092002, 2022.

\bibitem{Kaplan:2008ie}
David~E. Kaplan, Keith Rehermann, Matthew~D. Schwartz, and Brock Tweedie.
\newblock {Top Tagging: A Method for Identifying Boosted Hadronically Decaying
  Top Quarks}.
\newblock {\em Phys. Rev. Lett.}, 101:142001, 2008.

\bibitem{Cui:2010km}
Yanou Cui, Zhenyu Han, and Matthew~D. Schwartz.
\newblock {W-jet Tagging: Optimizing the Identification of Boosted
  Hadronically-Decaying W Bosons}.
\newblock {\em Phys. Rev. D}, 83:074023, 2011.

\bibitem{Plehn:2011sj}
Tilman Plehn, Michael Spannowsky, and Michihisa Takeuchi.
\newblock {How to Improve Top Tagging}.
\newblock {\em Phys. Rev. D}, 85:034029, 2012.

\bibitem{Soper:2012pb}
Davison~E. Soper and Michael Spannowsky.
\newblock {Finding top quarks with shower deconstruction}.
\newblock {\em Phys. Rev. D}, 87:054012, 2013.

\bibitem{Anders:2013oga}
Christoph Anders, Catherine Bernaciak, Gregor Kasieczka, Tilman Plehn, and
  Torben Schell.
\newblock {Benchmarking an even better top tagger algorithm}.
\newblock {\em Phys. Rev. D}, 89(7):074047, 2014.

\bibitem{Kasieczka:2015jma}
Gregor Kasieczka, Tilman Plehn, Torben Schell, Thomas Strebler, and Gavin~P.
  Salam.
\newblock {Resonance Searches with an Updated Top Tagger}.
\newblock {\em JHEP}, 06:203, 2015.

\bibitem{Thaler:2010tr}
Jesse Thaler and Ken Van~Tilburg.
\newblock {Identifying Boosted Objects with N-subjettiness}.
\newblock {\em JHEP}, 03:015, 2011.

\bibitem{Thaler:2011gf}
Jesse Thaler and Ken Van~Tilburg.
\newblock {Maximizing Boosted Top Identification by Minimizing N-subjettiness}.
\newblock {\em JHEP}, 02:093, 2012.

\bibitem{Larkoski:2013eya}
Andrew~J. Larkoski, Gavin~P. Salam, and Jesse Thaler.
\newblock {Energy Correlation Functions for Jet Substructure}.
\newblock {\em JHEP}, 06:108, 2013.

\bibitem{Moult:2016cvt}
Ian Moult, Lina Necib, and Jesse Thaler.
\newblock {New Angles on Energy Correlation Functions}.
\newblock {\em JHEP}, 12:153, 2016.

\bibitem{Larkoski:2014wba}
Andrew~J. Larkoski, Simone Marzani, Gregory Soyez, and Jesse Thaler.
\newblock {Soft Drop}.
\newblock {\em JHEP}, 05:146, 2014.

\bibitem{Abdesselam:2010pt}
A.~Abdesselam et~al.
\newblock {Boosted Objects: A Probe of Beyond the Standard Model Physics}.
\newblock {\em Eur. Phys. J. C}, 71:1661, 2011.

\bibitem{Altheimer:2012mn}
A.~Altheimer et~al.
\newblock {Jet Substructure at the Tevatron and LHC: New results, new tools,
  new benchmarks}.
\newblock {\em J. Phys. G}, 39:063001, 2012.

\bibitem{Altheimer:2013yza}
A.~Altheimer et~al.
\newblock {Boosted Objects and Jet Substructure at the LHC. Report of
  BOOST2012, held at IFIC Valencia, 23rd-27th of July 2012}.
\newblock {\em Eur. Phys. J. C}, 74(3):2792, 2014.

\bibitem{Cogan:2014oua}
Josh Cogan, Michael Kagan, Emanuel Strauss, and Ariel Schwarztman.
\newblock {Jet-Images: Computer Vision Inspired Techniques for Jet Tagging}.
\newblock {\em JHEP}, 02:118, 2015.

\bibitem{Almeida:2015jua}
Leandro~G. Almeida, Mihailo Backovi\'c, Mathieu Cliche, Seung~J. Lee, and Maxim
  Perelstein.
\newblock {Playing Tag with ANN: Boosted Top Identification with Pattern
  Recognition}.
\newblock {\em JHEP}, 07:086, 2015.

\bibitem{deOliveira:2015xxd}
Luke de~Oliveira, Michael Kagan, Lester Mackey, Benjamin Nachman, and Ariel
  Schwartzman.
\newblock {Jet-images \textemdash{} deep learning edition}.
\newblock {\em JHEP}, 07:069, 2016.

\bibitem{Baldi:2016fql}
Pierre Baldi, Kevin Bauer, Clara Eng, Peter Sadowski, and Daniel Whiteson.
\newblock {Jet Substructure Classification in High-Energy Physics with Deep
  Neural Networks}.
\newblock {\em Phys. Rev. D}, 93(9):094034, 2016.

\bibitem{Barnard:2016qma}
James Barnard, Edmund~Noel Dawe, Matthew~J. Dolan, and Nina Rajcic.
\newblock {Parton Shower Uncertainties in Jet Substructure Analyses with Deep
  Neural Networks}.
\newblock {\em Phys. Rev. D}, 95(1):014018, 2017.

\bibitem{Komiske:2016rsd}
Patrick~T. Komiske, Eric~M. Metodiev, and Matthew~D. Schwartz.
\newblock {Deep learning in color: towards automated quark/gluon jet
  discrimination}.
\newblock {\em JHEP}, 01:110, 2017.

\bibitem{Kasieczka:2017nvn}
Gregor Kasieczka, Tilman Plehn, Michael Russell, and Torben Schell.
\newblock {Deep-learning Top Taggers or The End of QCD?}
\newblock {\em JHEP}, 05:006, 2017.

\bibitem{Macaluso:2018tck}
Sebastian Macaluso and David Shih.
\newblock {Pulling Out All the Tops with Computer Vision and Deep Learning}.
\newblock {\em JHEP}, 10:121, 2018.

\bibitem{Choi:2018dag}
Suyong Choi, Seung~J. Lee, and Maxim Perelstein.
\newblock {Infrared Safety of a Neural-Net Top Tagging Algorithm}.
\newblock {\em JHEP}, 02:132, 2019.

\bibitem{Mokhtar:2022pwm}
Farouk Mokhtar, Raghav Kansal, and Javier Duarte.
\newblock {Do graph neural networks learn traditional jet substructure?}
\newblock In {\em {36th Conference on Neural Information Processing Systems}:
  {Workshop on Machine Learning and the Physical Sciences}}, 11 2022.

\bibitem{Ma:2022bvt}
Fei Ma, Feiyi Liu, and Wei Li.
\newblock {Jet tagging algorithm of graph network with Haar pooling message
  passing}.
\newblock {\em Phys. Rev. D}, 108(7):072007, 2023.

\bibitem{Gong:2022lye}
Shiqi Gong, Qi~Meng, Jue Zhang, Huilin Qu, Congqiao Li, Sitian Qian, Weitao Du,
  Zhi-Ming Ma, and Tie-Yan Liu.
\newblock {An efficient Lorentz equivariant graph neural network for jet
  tagging}.
\newblock {\em JHEP}, 07:030, 2022.

\bibitem{Guest:2016iqz}
Daniel Guest, Julian Collado, Pierre Baldi, Shih-Chieh Hsu, Gregor Urban, and
  Daniel Whiteson.
\newblock {Jet Flavor Classification in High-Energy Physics with Deep Neural
  Networks}.
\newblock {\em Phys. Rev. D}, 94(11):112002, 2016.

\bibitem{Pearkes:2017hku}
Jannicke Pearkes, Wojciech Fedorko, Alison Lister, and Colin Gay.
\newblock {Jet Constituents for Deep Neural Network Based Top Quark Tagging}.
\newblock 4 2017.

\bibitem{Egan:2017ojy}
Shannon Egan, Wojciech Fedorko, Alison Lister, Jannicke Pearkes, and Colin Gay.
\newblock {Long Short-Term Memory (LSTM) networks with jet constituents for
  boosted top tagging at the LHC}.
\newblock 11 2017.

\bibitem{Fraser:2018ieu}
Katherine Fraser and Matthew~D. Schwartz.
\newblock {Jet Charge and Machine Learning}.
\newblock {\em JHEP}, 10:093, 2018.

\bibitem{Butter:2017cot}
Anja Butter, Gregor Kasieczka, Tilman Plehn, and Michael Russell.
\newblock {Deep-learned Top Tagging with a Lorentz Layer}.
\newblock {\em SciPost Phys.}, 5(3):028, 2018.

\bibitem{Kasieczka:2018lwf}
Gregor Kasieczka, Nicholas Kiefer, Tilman Plehn, and Jennifer~M. Thompson.
\newblock {Quark-Gluon Tagging: Machine Learning vs Detector}.
\newblock {\em SciPost Phys.}, 6(6):069, 2019.

\bibitem{Alwall:2014hca}
J.~Alwall, R.~Frederix, S.~Frixione, V.~Hirschi, F.~Maltoni, O.~Mattelaer,
  H.~S. Shao, T.~Stelzer, P.~Torrielli, and M.~Zaro.
\newblock {The automated computation of tree-level and next-to-leading order
  differential cross sections, and their matching to parton shower
  simulations}.
\newblock {\em JHEP}, 07:079, 2014.

\bibitem{Porod:2003um}
Werner Porod.
\newblock {SPheno, a program for calculating supersymmetric spectra, SUSY
  particle decays and SUSY particle production at e+ e- colliders}.
\newblock {\em Comput. Phys. Commun.}, 153:275--315, 2003.

\bibitem{Porod:2011nf}
W.~Porod and F.~Staub.
\newblock {SPheno 3.1: Extensions including flavour, CP-phases and models
  beyond the MSSM}.
\newblock {\em Comput. Phys. Commun.}, 183:2458--2469, 2012.

\bibitem{Sjostrand:2006za}
Torbjorn Sjostrand, Stephen Mrenna, and Peter~Z. Skands.
\newblock {PYTHIA 6.4 Physics and Manual}.
\newblock {\em JHEP}, 05:026, 2006.

\bibitem{deFavereau:2013fsa}
J.~de~Favereau, C.~Delaere, P.~Demin, A.~Giammanco, V.~Lema\^\i{}tre,
  A.~Mertens, and M.~Selvaggi.
\newblock {DELPHES 3, A modular framework for fast simulation of a generic
  collider experiment}.
\newblock {\em JHEP}, 02:057, 2014.

\bibitem{Alwall:2007fs}
Johan Alwall et~al.
\newblock {Comparative study of various algorithms for the merging of parton
  showers and matrix elements in hadronic collisions}.
\newblock {\em Eur. Phys. J. C}, 53:473--500, 2008.

\bibitem{Mangano:2006rw}
Michelangelo~L. Mangano, Mauro Moretti, Fulvio Piccinini, and Michele Treccani.
\newblock {Matching matrix elements and shower evolution for top-quark
  production in hadronic collisions}.
\newblock {\em JHEP}, 01:013, 2007.

\bibitem{Cacciari:2008gp}
Matteo Cacciari, Gavin~P. Salam, and Gregory Soyez.
\newblock {The anti-$k_t$ jet clustering algorithm}.
\newblock {\em JHEP}, 04:063, 2008.

\bibitem{Catani:1993hr}
S.~Catani, Yuri~L. Dokshitzer, M.~H. Seymour, and B.~R. Webber.
\newblock {Longitudinally invariant $K_t$ clustering algorithms for hadron
  hadron collisions}.
\newblock {\em Nucl. Phys. B}, 406:187--224, 1993.

\bibitem{Krohn:2009th}
David Krohn, Jesse Thaler, and Lian-Tao Wang.
\newblock {Jet Trimming}.
\newblock {\em JHEP}, 02:084, 2010.

\bibitem{Cowan:2010js}
Glen Cowan, Kyle Cranmer, Eilam Gross, and Ofer Vitells.
\newblock {Asymptotic formulae for likelihood-based tests of new physics}.
\newblock {\em Eur. Phys. J. C}, 71:1554, 2011.
\newblock [Erratum: Eur.Phys.J.C 73, 2501 (2013)].

\bibitem{LHCDarkMatterWorkingGroup:2018ufk}
Tomohiro Abe et~al.
\newblock {LHC Dark Matter Working Group: Next-generation spin-0 dark matter
  models}.
\newblock {\em Phys. Dark Univ.}, 27:100351, 2020.

\bibitem{Arganda:2022idg}
Ernesto Arganda, Antonio Delgado, Roberto~A. Morales, and Mariano Quir\'os.
\newblock {LHC Search Strategy for Squarks in Higgsino-LSP Scenarios with
  Leptons and b-Jets in the Final State}.
\newblock {\em Particles}, 5(3):265--272, 2022.

\bibitem{selvaraju2017grad}
Ramprasaath~R Selvaraju, Michael Cogswell, Abhishek Das, Ramakrishna Vedantam,
  Devi Parikh, and Dhruv Batra.
\newblock Grad-cam: Visual explanations from deep networks via gradient-based
  localization.
\newblock In {\em Proceedings of the IEEE international conference on computer
  vision}, pages 618--626, 2017.

\bibitem{huang2022ssit}
Yijin Huang, Junyan Lyu, Pujin Cheng, Roger Tam, and Xiaoying Tang.
\newblock Ssit: Saliency-guided self-supervised image transformer for diabetic
  retinopathy grading.
\newblock {\em arXiv preprint arXiv:2210.10969}, 2022.

\bibitem{duong2022put}
Nghia Duong-Trung, Stefan Born, Kiran Madhusudhanan, Randolf Scholz, Johannes
  Burchert, Danh Le-Phuoc, and Lars Schmidt-Thieme.
\newblock Put attention to temporal saliency patterns of multi-horizon time
  series.
\newblock {\em arXiv preprint arXiv:2212.07771}, 2022.

\bibitem{lu2023saliency}
Changsheng Lu, Hao Zhu, and Piotr Koniusz.
\newblock From saliency to dino: Saliency-guided vision transformer for
  few-shot keypoint detection.
\newblock {\em arXiv preprint arXiv:2304.03140}, 2023.

\bibitem{binder2016layer}
Alexander Binder, Gr{\'e}goire Montavon, Sebastian Lapuschkin, Klaus-Robert
  M{\"u}ller, and Wojciech Samek.
\newblock Layer-wise relevance propagation for neural networks with local
  renormalization layers.
\newblock In {\em Artificial Neural Networks and Machine Learning--ICANN 2016:
  25th International Conference on Artificial Neural Networks, Barcelona,
  Spain, September 6-9, 2016, Proceedings, Part II 25}, pages 63--71. Springer,
  2016.

\bibitem{cherepanov2022visualization}
Igor Cherepanov, Alex Ulmer, Jonathan~Geraldi Joewono, and J{\"o}rn Kohlhammer.
\newblock Visualization of class activation maps to explain ai classification
  of network packet captures.
\newblock In {\em 2022 IEEE Symposium on Visualization for Cyber Security
  (VizSec)}, pages 1--11. IEEE, 2022.

\bibitem{zhou2016learning}
Bolei Zhou, Aditya Khosla, Agata Lapedriza, Aude Oliva, and Antonio Torralba.
\newblock Learning deep features for discriminative localization.
\newblock In {\em Proceedings of the IEEE conference on computer vision and
  pattern recognition}, pages 2921--2929, 2016.

\bibitem{chefer2021transformer}
Hila Chefer, Shir Gur, and Lior Wolf.
\newblock Transformer interpretability beyond attention visualization.
\newblock In {\em Proceedings of the IEEE/CVF conference on computer vision and
  pattern recognition}, pages 782--791, 2021.

\bibitem{kingma2014adam}
Diederik~P Kingma and Jimmy Ba.
\newblock Adam: A method for stochastic optimization.
\newblock {\em arXiv preprint arXiv:1412.6980}, 2014.

\bibitem{terven2023loss}
Juan Terven, Diana~M Cordova-Esparza, Alfonzo Ramirez-Pedraza, and Edgar~A
  Chavez-Urbiola.
\newblock Loss functions and metrics in deep learning. a review.
\newblock {\em arXiv preprint arXiv:2307.02694}, 2023.

\bibitem{lai2022rethinking}
Zhitong Lai, Haichao Sun, Rui Tian, Nannan Ding, Zhiguo Wu, and Yanjie Wang.
\newblock Rethinking skip connections in encoder-decoder networks for monocular
  depth estimation.
\newblock {\em arXiv preprint arXiv:2208.13441}, 2022.

\bibitem{hendrycks2016gaussian}
Dan Hendrycks and Kevin Gimpel.
\newblock Gaussian error linear units (gelus).
\newblock {\em arXiv preprint arXiv:1606.08415}, 2016.

\end{thebibliography}
%%%%%%%%%%%%%%%%%%%%
\end{document}